\documentclass[11pt,a4paper,openright]{article}
\usepackage{jheppub}

\usepackage{braket}
\usepackage{multirow}
\usepackage{morefloats}
\usepackage{pb-diagram}
\usepackage{tikz}
\usetikzlibrary{matrix,snakes,arrows,shapes,decorations.pathmorphing,decorations.markings,calc}

\title{The deconfining phase transition of SO(N) gauge theories in 2+1 dimensions}
\author{Richard Lau,}
\author{Michael Teper}
\affiliation{Rudolf Peierls Centre for Theoretical Physics,
  \\University of Oxford, \\1 Keble Road, \\Oxford, OX1 3NP, UK}
\emailAdd{richard.lau@physics.ox.ac.uk}
\emailAdd{m.teper1@physics.ox.ac.uk}
\abstract{We calculate the deconfining temperature of $SO(N)$ gauge theories
  in 2+1 dimensions, and determine the order of the phase transition as a
  function of $N$, for various values of $N\in [4,16]$.  
  We do so by extrapolating our lattice results to the infinite volume limit,
  and then to the continuum limit, for each value of $N$. We then extrapolate
  to the $N=\infty$ limit and observe that the $SO(N)$ and $SU(N)$ deconfining
  temperatures agree in that limit.
  We find that the the deconfining temperatures of all the $SO(N)$ gauge 
  theories appear to follow a single smooth function of $N$, despite
  the lack of a non-trivial centre for odd $N$.
  We also compare the deconfining temperatures of $SO(6)$ with $SU(4)$, and
  of $SO(4)$ with $SU(2) \times SU(2)$, motivated by the fact that these pairs
  of gauge theories share the same Lie algebras.}

\begin{document}
\maketitle

\section{Introduction}
\label{sec:introduction}

While a great deal is known about the non-perturbative physics of $SU(N)$ gauge theories 
from calculations on the lattice, much less is known about $SO(N)$ gauge theories. 
In this paper we will show that $SO(N)$ gauge theories in $2+1$ dimensions possess
a deconfining phase transition at a finite temperature $T=T_c$, just like the deconfining 
transition in $SU(N)$ gauge theories. We will calculate its value and  determine its nature for  
$N=4,5,6,7,8,9,12,16$. This will enable us to extrapolate to $N=\infty$ where we can 
compare to the $SU(\infty)$ extrapolated value
\cite{Jack-Liddle:2008xr}. 
This is interesting to do since  $SO(N)$ and $SU(N)$ gauge theories have a common planar limit
\cite{Lovelace:1982fk},
and $SO(2N)$ and $SU(N)$ gauge theories are orbifold equivalent
\cite{orbifold},
so we expect that dimensionless ratios of common physical quantities, including the 
deconfining temperature, should be equal at $N=\infty$
\cite{Unsal:2006kx}. 
We will perform further comparisons motivated by the fact that certain $SO(N)$ and  
$SU(N^\prime)$ gauge theories share 
the same Lie algebras, i.e. $SO(3)$ and $SU(2)$, $SO(4)$ and $SU(2)\times SU(2)$, 
$SO(6)$ and $SU(4)$. To the extent that the difference in the global properties of 
the groups (such as the centre) is not important, we would expect the deconfining 
transition and temperature to be identical within each of these pairs of gauge 
theories, and this is something we shall attempt to check. Moreover assuming this 
identity, the known value of $T_c$ in $SU(2)$ provides us with a value for $SO(3)$,
which we do not calculate directly (for reasons given below).  
In addition all these calculations will allow us to compare $SO(2N)$ and $SO(2N+1)$ theories, 
which is interesting because $SO(2N+1)$ gauge theories have a trivial center in contrast 
to the non-trivial $\mathbb{Z}_2$ center of $SO(2N)$ theories.

While the calculations in this paper are primarily intended to establish the presence of 
the finite $T$ transition and to investigate its properties, we shall choose to call it
a deconfining transition, for both odd and even $N$, just like  the one in $SU(N)$ 
gauge theories. Of course that assumes that these theories are linearly confining 
at low $T$. While  we shall provide some evidence for confinement at low $T$ in this 
paper (see in particular the discussion in Section~\ref{subsec:poly}), the explicit 
evidence for the confinement being linear is given in our companion paper on the 
glueball spectra and string tensions
\cite{Lau:aa},
where we show that the energy of closed flux tubes is (roughly) proportional to
their length for both odd and even $N$. Of course such numerical evidence possesses  
intrinsic limitations: we cannot distinguish between confinement that is exact
and confinement to a very good approximation. However the quality of our numerical
evidence is comparable to that which establishes linear confinement in $D=2+1$ $SU(N)$ 
gauge theories.

The paper is structured as follows.
In Section~\ref{sec:equivalences}, we briefly review some well-known relations 
between $SO(N)$ and $SU(N)$ gauge theories, both at small and at large $N$.
In Section~\ref{sec:preliminaries} we briefly describe the lattice setup, 
how to differentiate confining from non-confining phases, and we comment
on what we know about confinement in $SO(N)$ gauge theories.
In Section~\ref{sec:deconfinement}, we describe how we identify the location 
of the finite temperature transition and how we determine whether the transition is 
first or second order.
Then in Section~\ref{sec:lattice}, we describe how to calculate on a  lattice
the physical quantities that we shall use in order to express the transition 
temperature in physical units.
The next few sections contain our results.
First, in Section~\ref{sec:infinitevolume} , we calculate the infinite volume limit 
for each of the $SO(N)$ gauge theories we consider, and hence the value of $T_c$ at 
various lattice spacings. 
Then, in Section~\ref{sec:continuum}, we use these values to calculate the continuum 
limit of the deconfining temperature for each group,
briefly discussing the issues caused by the strong to weak coupling `bulk' transition 
in $D=2+1$.
We then proceed in Section~\ref{sec:largen} to calculate the large-$N$ limit of 
$T_c$ for $SO(2N)$ and $SO(2N+1)$ separately and together and
in Section~\ref{sec:comparison}, we compare the $SO(N)$ and $SU(N)$ deconfining 
temperatures both at $N=\infty$, and for pairs of  $SO(N)$ and $SU(N^\prime)$ 
groups that share the same Lie algebra.
Section~\ref{sec:conclusions}, contains a summary of our conclusions.
Appendix~\ref{app:data} contains our detailed tabulated results.

There are companion papers, both published 
\cite{Athenodorou:2015aa}
and in progress
\cite{Lau:aa},
that contain our results for the mass spectrum and string tension of $SO(N)$ gauge 
theories. The latter paper describes the Monte Carlo algorithm in more detail, as 
well as providing more discussion of the `bulk' transition. An earlier paper 
\cite{Bursa:2013lq}
contained our first, exploratory estimates of $T_c$. The values in the present paper 
are much more accurate and supersede those earlier values, although they are in 
fact consistent within errors.

\section{Relations between $SO(N)$ and $SU(N)$}
\label{sec:equivalences}

\subsection{Lie algebra equivalences}
\label{subsec:Lie-equivalences}

The implications of the Lie algebra equivalence between certain $SO(N)$ and
$SU(N^\prime)$ groups are discussed in more detail in
\cite{Athenodorou:2015aa,Lau:aa}.
Here we merely summarise some points that are relevant to our present calculations. 
If we assume that the global structure of the groups is irrelevant to the physics 
(an assumption which needs to be tested) then we expect that colour singlet quantities 
are the same within each pair of theories. For example the mass gap or the deconfining 
temperature. String tensions on the other hand are associated with a flux that is in 
a certain representation, and this needs to be matched between the theories. In the 
following we summarise some points that are relevant to the calculations in this paper.

The first equivalence is between the Lie algebras of $SO(3)$ and $SU(2)$. The $SO(3)$ 
fundamental representation is equivalent to the $SU(2)$ adjoint representation, so 
that the associated string tension should satisfy
\begin{equation}
\left.\sigma_f\right|_{so3} = \left.\sigma_{adj}\right|_{su2}.
\label{}
\end{equation}
Adjoint flux tubes in $SU(2)$ are not expected to be stable and can, for example, 
decay into glueballs. So one expects the same to be true for $SO(3)$ fundamental 
flux tubes. Of course, if the decay width is small enough, then just as for the 
mass of a narrow resonance, we can estimate a string tension. More importantly, 
glueball masses and $T_c$ should be the same within $SU(2)$ and $SO(3)$, and the 
coupling $g^2$ should satisfy 
\cite{Bursa:2013lq,Lau:aa}
\begin{equation}
\left.g^2\right|_{so3} = \left.4g^2\right|_{su2}.
\label{}
\end{equation} 
In this paper we do not calculate $T_c$ for $SO(3)$ because the strong-to-weak 
coupling transition in our lattice theory occurs at such a small value of the 
lattice spacing that we would need to use very large lattices (in lattice units) 
and this would be computationally quite expensive. Of course, if one assumes 
that the physics of $SU(2)$ and $SO(3)$ is the same, as described above, then 
one can infer the value of $T_c$ in $SO(3)$ from the known value in $SU(2)$, 
and compare it to the values obtained in $SO(N > 3)$. We shall do this later in this paper.

As is also well known, $SO(4)$ and $SU(2)\times SU(2)$ share the same Lie algebra, 
with the latter forming a double cover of the former. In an  $SU(2)\times SU(2)$ 
theory the two $SU(2)$ groups do not interact with each other and so the physics 
is directly related to that of $SU(2)$. So if we assume that the physics of the  
$SO(4)$ and $SU(2)\times SU(2)$ gauge  theories is the same, then the single particle 
spectrum and the value of $T_c$ should be just as in $SU(2)$. Because the fundamental 
$SO(4)$ flux involves fundamental flux from both $SU(2)$ groups, we expect     
\begin{equation}
\left.\sigma_f\right|_{so4} = \left. 2 \sigma_f\right|_{su2}.
\label{}
\end{equation}
We also expect the couplings to be related by
\cite{Bursa:2013lq,Lau:aa}
\begin{equation}
\left.g^2\right|_{so4} = \left.2g^2\right|_{su2}.
\label{}
\end{equation}

Finally, we recall that $SO(6)$ and $SU(4)$ also share the same Lie algebra. 
We further recall that in $SU(4)$ 
\begin{align}
\underline{4}\otimes\underline{4}=\underline{6}\oplus\underline{10}
\end{align}
where the $\underline{6}$ corresponds to the $k=2$ antisymmetric representation and this maps 
to the fundamental $\underline{6}$ of $SO(6)$.
To convert quantities in terms of the $SU(4)$ fundamental string tension to the 
$SU(4)$ $k=2A$ string tension, we shall use the known ratio of the $SU(4)$ $k=2A$ 
and fundamental string tensions in $D=2+1$
\cite{Bringoltz2008429} 
\begin{align}
\left.\frac{\sigma_{2A}}{\sigma_{f}}\right|_{su4}=1.355(9).
\end{align}
In addition the couplings are related by
\cite{Bursa:2013lq,Lau:aa}
\begin{equation}
\left.g^2\right|_{so6} = \left.2g^2\right|_{su4}.
\label{}
\end{equation} 
Glueball masses and $T_c$ should be the same for $SO(6)$ and $SU(4)$, in their 
common positive charge conjugation sector. 

All the above relations assume that the differing global properties of the 
pairs of gauge groups do not affect the dynamics. It is not obvious that  
this is the case and one of our aims in this paper is to see if it is indeed
the case for the properties of the deconfining transition.

\subsection{Large-$N$}
\label{subsec:largeN-equivalences}

Just as with $SU(N)$ gauge theories 
\cite{Hooft1974461}, 
$SO(N)$ gauge theories at the diagrammatic level possess a smooth $N\to\infty$ limit 
if one keeps $g^2N$ fixed
\cite{Lovelace:1982fk}.
Moreover the surviving planar diagrams are identical to those of $SU(N)$ if one chooses
\cite{Lovelace:1982fk}
\begin{equation}
\left.g^2\right|_{SO(N)} = \left.2g^2\right|_{SU(N)}.
\label{gsonsun}
\end{equation} 
However there is a difference in the approach to the planar limit. The $SO(N)$ gauge 
field propagator takes the form 
\begin{align}
\Braket{[A_\mu(x)]^{i}{}_j [A_\nu(y)]^{k}{}_l} 
\propto \delta^{i}{}_l\delta^{k}{}_j-\delta^{ik}\delta_{lj} .
\end{align} 
The first term on the right is the leading order double line description of an $SU(N)$ 
gauge propagator. However, the second term is special to $SO(N)$ gauge theories and 
corresponds to a `twisted' propagator 
\cite{Blake:2012kx}. 
This leads to  new non-oriented surfaces in double line graphs, which in turn means 
that corrections to the planar limit are $O(1/N)$ rather than the $O(1/N^2)$
one finds for $SU(N)$.

While the above diagrammatic analysis suggests that the large-$N$ physics of $SU(N)$ 
and $SO(N)$ gauge theories should be the same in their common positive charge
conjugation sector of states, it does 
not guarantee that non-perturbative effects will not disrupt this expectation. However 
there exists a more general argument based on a large-$N$ orbifold equivalence
\cite{orbifold}.
One can apply an orbifold projection on a parent $SO(2N)$ QCD-like theory to obtain a 
child $SU(N)$ QCD theory
\cite{orbifold}
with the couplings related as in eqn(\ref{gsonsun}). Since it has been shown that the
large-$N$ physics of orbifold equivalent theories is indeed the same
\cite{Unsal:2006kx},
this tells us that the physics of large-$N$ $SO(2N)$ and $SU(N)$ gauge theories should be 
identical within their common sector. In particular this should apply to the $N\to\infty$
limit of the calculations of $T_c$ in this paper.

\section{Preliminaries}
\label{sec:preliminaries}

\subsection{Lattice variables}

Our variables are $N\times N$ $SO(N)$ matrices $U_l$ assigned to links $l$. We will often 
write $U_l$ as $U_\mu(x)$ where the link $l$ emanates from the site $x$ in the 
$\mu$ direction. Our periodic lattice has dimensions  $L_s^2L_t$, with lattice 
spacing $a$. The partition function is
\begin{align}
  Z & =  \int \prod_l dU_l \exp \{-\beta S[U_l] \}
\end{align}
and we use a standard plaquette action $\beta S$ where
\begin{align}
\beta S & =  \beta\sum_{p}\left( 1-\frac{1}{N} \text{Tr}(U_p) \right)  \nonumber\\
\beta	& =\frac{2N}{ag^2} 
\label{eqn:plaqaction}
\end{align}
with $U_p$ the ordered product of $U_l$ around the boundary of the plaquette $p$.
The relation between $\beta$ and $g^2$ holds in the continuum limit; on the
lattice it defines a lattice coupling that will become the standard continuum
coupling when $a \rightarrow 0$.
Note that different choices of action will lead to definitions of $g^2$ that 
differ by $O(a)$ corrections (which of course vanish in the continuum limit).

\subsection{Finite temperature on the lattice}

To calculate expectation values at a non-zero temperature $T$, we consider the
Euclidean field theory on a periodic $l^2_sl_t$ space-time volume and take the
thermodynamic limit $l_s\to\infty$ so that we have a well-defined temperature,
$l_t=1/T$.

For convenience we shall use $T=1/l_t$ to define the `temperature' of our
system even in a finite volume.

On a $L_s^2L_t$ lattice with spacing $a$, we have $l_s=aL_s$ and $l_t=aL_t$.
The value of $a$ is determined by the value of the inverse bare coupling,
$\beta = 2N/ag^2$, that appears in the lattice action. So for $L_s\to\infty$
a lattice field theory will have temperature $T=1/a(\beta)L_t$.
We can vary $T$ at fixed $L_t$ by varying $\beta$ and hence $a(\beta)$.
If we find that a deconfinement transition occurs at $\beta=\beta_c$, then
the deconfining temperature is
\begin{align}
T_c(a)=\frac{1}{a(\beta_c)L_t}.
\end{align}
If we increase $L_t$, the transition will occur at a smaller value of $a$.
So by producing a sequence of such calculations we can extrapolate to the
$a=0$ continuum limit.

\subsection{The `temporal' Polyakov loop,  the center,  and confinement}
\label{subsec:poly}

A useful order parameter for identifying the deconfining transition is the 
`temporal' Polyakov loop, $l_P$. If the spatial starting point of the loop 
is $\mathbf{x}$, then the loop is defined by
 \begin{align}
l_P(\mathbf{x}) 	
&=\text{Tr}\left(U_t(\mathbf{x},t=a)U_t(\mathbf{x},t=2a)\cdots 
U_t(\mathbf{x},t=aL_t)\right). 
\label{phase:initialpolyakovloop}
\end{align}
This operator represents the world line of a static charge in the fundamental 
representation located at spatial site $\mathbf{x}$. So we can obtain the free 
energy $F_{f\bar{f}}$ of a pair of such charges located at $\mathbf{x}$ and 
$\mathbf{y}$ respectively from the correlation function of two Polyakov loops
at $\mathbf{x}$ and $\mathbf{y}$ with opposite orientations
\begin{align}
e^{-\tfrac{1}{T}F_{f\bar{f}}(x,y)}=\braket{l_P(\mathbf{x})l_P^T(\mathbf{y})}.
\end{align}
Assuming that the correlation function satisfies clustering, the correlation 
function decorrelates at large spatial distances
\begin{align}
\braket{l_P(\mathbf{x})l_P^T(\mathbf{y})}\xrightarrow[\left|\mathbf{x}
-\mathbf{y}\right|\rightarrow\infty]{} \lvert \braket{l_P}\rvert^2.
\end{align}
Hence, if $\braket{l_P}=0$ then $F_{f\bar{f}}(x,y) \to \infty$ as the separation 
$\left|\mathbf{x}-\mathbf{y}\right|\rightarrow\infty$ which corresponds to confinement, 
although not necessarily to a linearly rising potential. (Recall that in $D=2+1$ 
the Coulomb interaction is already, by itself, logarithmically confining.)
Similarly, if $\braket{l_P}\neq0$, then the free energy approaches a finite value 
at large spatial separation, and this will normally imply that the charges are not 
confined. A counterexample is when there are particles in the fundamental representation 
in the theory, which can then bind with the static charge to produce a colour singlet.
This is the case in $QCD$ where
we have  $\braket{l_P}\neq0$, but the theory is confining (all physical states
are colour singlet) even though the potential flattens out at large distances.

$SO(2N)$ gauge theories have a $\mathbb{Z}_2$ centre symmetry under which the action and 
measure are invariant. We can generate a centre symmetry transformation by taking 
a non-trivial element  $z$ of the centre and multiplying all temporal links 
between two neighbouring  time-slices by $z$. Unlike a contractible loop, the
temporal Polyakov loop is not invariant under this symmetry, 
\begin{align}
l_P\rightarrow zl_P
\end{align}
so that its expectation value $\braket{l_P}=0$, and the theory is confining, unless 
the centre symmetry is spontaneously broken, in which case we generically expect 
$\braket{l_P}\neq 0$ and the theory is deconfining. So we expect that the 
deconfinement phase transition coincides with the spontaneous breakdown of the 
centre symmetry. This, of course, just parallels the well-known argument for 
$SU(N)$ gauge theories. In addition the Lie algebra equivalences discussed
in Section~\ref{subsec:Lie-equivalences} strongly suggest that both $SO(4)$ and
$SO(6)$ must be confining at low $T$, just like $SU(2)$ and $SU(4)$ respectively.
Moreover the large-$N$ equivalences discussed in  
Section~\ref{subsec:largeN-equivalences} strongly suggest that the $SO(N\to\infty)$
theory is linearly confining, just like  $SU(N\to\infty)$. All this (together
with the numerical evidence for linear confinement in 
\cite{Lau:aa}),
makes a convincing case that $SO(2N)$ gauge theories are linearly confining at
low $T$.

By contrast $SO(2N+1)$ gauge theories have a trivial centre  and so in general 
we would expect $\braket{l_P}\neq0$ at all $T$. Even so, this does not 
of itself preclude confinement. As a well-known example, recall that in 
$QCD$ with a heavy enough but finite quark mass one has $\braket{l_P}\neq0$, 
albeit very small, because of the explicit breaking of the centre symmetry by 
the fermion action. Nonetheless the theory still possesses a first order 
deconfining transition, which is continuously linked to that 
of the pure gauge theory (which is why we confidently label it as being
deconfining). In $QCD$ in this limit a long confining flux tube 
is in fact unstable, but with an extremely small decay width -- the breaking 
is essentially a tunnelling phenomenon. So strictly speaking the theory is 
not linearly confining, although it is still believed to be physically 
confining in the sense that all finite-energy states are colour singlet.
(And in practice the flux-tube breaking  would not be visible in a direct numerical 
calculation of the potential.) Another well-known and more relevant
example is provided by the $SO(3)$ gauge theory. The Lie algebra 
equivalence with $SU(2)$ (see Section~\ref{subsec:Lie-equivalences})  
strongly suggests that $SO(3)$ is confining 
at low $T$ with a second order deconfining transition at some non-zero $T$. 
However the fundamental flux tube of $SO(3)$ is the adjoint flux tube of 
$SU(2)$ which we expect to be unstable so that in $SO(3)$ we are confident that
$\braket{l_P}\neq0$ at any $T$. Indeed the direct physical interpretation of this 
is that the $SO(3)$ fundamental source is screened by  gluons which, in $SO(3)$, 
are in the same triplet as the fundamental. (Something that is not the case 
for $SO(N\geq 4)$.) A further directly relevant example is provided
by $SO(5)$. This has the same Lie algebra as $Sp(2)$. (Note that there
is another convention where this is called $Sp(4)$.) There have been
numerical investigations of $D=2+1$ $Sp(2)$ demonstrating that it 
has a second order deconfining transition
\cite{Pepe:2004}. 
So we strongly expect $SO(5)$ to also possess a deconfining transition.
Yet another useful example is provided by $G(2)$ which has a trivial center
and yet has a deconfining transition
\cite{Pepe:2006} 
from a low $T$ confining phase
\cite{Pepe:2003}.
(For a discussion of the centre and confinement see e.g
\cite{Pepe:2003}.)
Finally, the diagrammatic (not orbifold) equivalence 
between $SO(2N+1\to\infty)$ and $SU(N\to\infty)$ 
(see Section~\ref{subsec:largeN-equivalences}), strongly suggests that
in $SO(2N+1\to\infty)$ at $N=\infty$ we have exact confinement at $T=0$ and we 
also have $\braket{l_P}= 0$ at any $T$. Now if $SO(3)$ (not to mention $SO(5)$) 
and $SO(2N+1\to\infty)$ are exactly confining at low $T$, then it appears 
very plausible that all $SO(2N+1)$ gauge theories are exactly confining 
at low $T$. 

Even if $SO(2N+1)$ gauge theories are indeed confining, as we argued above, it is 
still interesting to ask if there is some exact order parameter based  on 
the Polyakov loop. Since $\braket{l_P}\neq0$ in $SO(3)$, but $\braket{l_P}=0$
at $N=\infty$, it is plausible that  $\braket{l_P}\neq0$ for any 
$SO(2N+1)$, but $\to 0$ as $N\to\infty$, and perhaps does this so
rapidly that the non-zero value becomes invisible in a numerical 
calculation at moderate values of (odd) $N$. Returning to $SO(3)$ we
observe that it is the fundamental Polyakov loop of $SU(2)$ that is exactly
zero at low $T$, and since this corresponds to the spinorial of $SO(3)$,
we expect that the corresponding spinorial Polyakov loop is exactly zero
in $SO(3)$. This suggests the following speculation. In $SO(2N+1)$ 
gauge theories it is perhaps the spinorial Polyakov loop that is exactly zero
(perhaps one can even locate a symmetry that ensures this) and this serves as 
the `ideal' order parameter for (de)confinement. But since the dimension of 
the spinorial representation in $SO(N)$ grows very rapidly with $N$, and 
one's experience is that string tensions grow very roughly with the 
quadratic Casimir, it will presumably only be relevant to the low energy 
physics at small $N$. Simultaneously, we expect that the expectation value 
of the fundamental loop in $SO(2N+1)$ decreases very rapidly, perhaps exponentially 
in $N$ if the tunnelling argument is correct, and it takes over as the `ideal' 
order parameter at larger $N$. Assessing the plausibility of such a scenario 
is something that we will not do here, or in
\cite{Lau:aa},
since it would require explicit calculations with the spinorial 
representations of $SO(N)$ gauge theories. But it is clearly something
that would be interesting to do.

A final practical comment. Later on in this paper we shall take $SO(7)$ 
as our typical example of $SO(2N+1)$ gauge theories, and we shall show that
the value of $\braket{l_P}$ at low $T$ is extremely small, and indeed 
consistent with zero within our very small errors. So we can assert that, at 
the very least, we have a direct numerical demonstration of something
close to exact confinement. And in
\cite{Lau:aa}
we shall show that, again within very small errors, this apparent confinement 
is in fact linear. Together with the above arguments this provides a justification
for labelling the finite $T$ transition that we study in this paper as being 
a `deconfining' one.

\section{Deconfining phase transitions}
\label{sec:deconfinement}

In an infinite spatial volume, a phase transition occurs when the free energy
becomes a non-analytic function in one of its parameters. 
We will see that the $SO(N)$ deconfining phase transition is second order
for small $N$ and first order for larger $N$.
First order phase transitions occur when there is a discontinuity in the
first derivative of the free energy such that the second derivative is
typically a delta function singularity. 
Second order phase transitions occur when there is a divergence in the
second derivative in the free energy although the first derivative is
continuous. This corresponds to a divergent correlation length.

On a finite volume, the partition function is finite so all derivatives
are well-defined and analytic, so that there are no apparent non-analyticities. 
Finite size scaling tells us how the results at finite volumes
should converge towards the expected non-analyticity as we increase the
spatial volume size, allowing us to classify the transition. 

\subsection{First order transitions}

Let $O$ be an order parameter, such as the temporal Polyakov loop or plaquette
averaged over the spatial volume.
Suppose  that it takes a value $\braket{O}=O_c$ in the confined phase and
$\braket{O}=O_d$ in the deconfined phase.
(For simplicity we shall assume here a single deconfined phase.)
We can define a susceptibility $\chi_O(V,T)$ for a volume $V$ and temperature $T$ by
\begin{align}
\chi_O(V,T)=\mathcal{N}V\left(\braket{O(T)^2}-\braket{O(T)}^2\right)
\end{align}
for some constant $\mathcal{N}$. If we are in a single phase then the spatial
average ensures that $\left(\braket{O(T)^2}-\braket{O(T)}^2\right) \sim O(1/V)$
so that $\chi_O(V,T)  \sim O(V^0)$, as long as the correlation length is finite,
i.e. the mass gap is non-zero.

At the phase transition, $T=T_c$, in an infinite volume the free energies are equal.
On a finite volume the phase transition is smeared out and there is no unique way 
to say at which value of $T$ it occurs, but a sensible and standard choice is to choose
$T_c$ where the free energy densities are equal
\begin{align}
f_c(T=T_c)=f_d(T=T_c)
\end{align}
where $F_{c/d}(T)=f_{c/d}(T)V$ are the free energies for the confined and deconfined
phases respectively.
At $T=T_c$ the system is equally likely to be in the confined and deconfined phases
and so the order parameter takes values $O_c$ and $O_d$ with equal probability. Hence,
\begin{align}
\chi_O(V,T_c)	&=\mathcal{N}V\left(\frac{(O_c^2+O_d^2)}{2}
-\frac{(O_c+O_d)^2}{4}\right)=\mathcal{N}V\left(\frac{(O_c-O_d)^2}{4}\right)
\end{align}
and so the peak height of the susceptibility should grow as $\chi_{\text{max}}=\mathcal{O}(V)$. 
Note that the susceptibility peaks when the probability of being in the confining phase
is $1/2$ and that this is independent of the number of identical deconfined phases.
Note also that here we neglect the $O(1/\surd V)$ fluctuations of $O$ around its
mean value in each phase.

So we conclude that a first order transition on finite volumes $V$ is characterised
by a susceptibility that forms a peak with height $\chi_{\text{max}}=\mathcal{O}(V)$ and 
that the whole peak is confined to a range
$\Delta \beta=\mathcal{O}(1/V)$. So as $V\to\infty$ the peak tends towards
a $\delta$-function and in extrapolating $T_c(V)$ to $V=\infty$ one should use 
a leading $\mathcal{O}(1/V)$ correction term.

\subsection{Second order transitions}

For a second order phase transition, the correlation length $\xi\to\infty$
as  $T\to T_c$ if we are on an infinite volume. On a finite volume it will
(effectively) approach the spatial lattice length $L_s$
\cite{Privman:ai}.
Let us define the reduced temperature by
\begin{align}
t&=(T-T_c)/T_c=(\beta-\beta_c)/\beta_c\equiv\Delta\beta
\end{align}
using $T=1/(aL_t)=\beta g^2/(2NL_t)$, and the critical exponents $\nu$ and 
$\gamma$ by the standard relations
\begin{align}
  \xi
  \sim\lvert t\rvert^{-\nu}
  \sim\lvert \Delta\beta\rvert^{-\nu}\nonumber\\
  \chi(T,L_s\rightarrow\infty)
  \sim\lvert t\rvert^{-\gamma}\sim\lvert \Delta\beta\rvert^{-\gamma}.
\end{align}
The standard finite size scaling analysis 
\cite{Privman:ai}
then tells us that at the transition the susceptibility has a 
height $\chi_{\text{max}}=\mathcal{O}(L_s^{\frac{\gamma}{\nu}})$
over a half-width of $\Delta\beta=\mathcal{O}(1/L_s^{\frac{1}{\nu}})$. 
Note that the $L_s\to\infty$ peak provides an envelope for the
peaks at finite $L_s$, leading to a structure quite different from the
$\delta$-function peak in a first order transition.

\subsection{Scaling laws}

From the above we infer that we can distinguish between first and second order 
transitions by examining the structure of the susceptibility peaks over a range 
of different spatial volumes.
We summarise the scaling laws by the following relations. 
In $D=2+1$, the phase transition occurs at
\begin{align}
  \frac{T_c(\infty)-T_c(V)}{T_c(\infty)}&\sim\frac{1}{V}
  &&\Rightarrow&&\beta_c(V)=\beta_c(\infty)\left[1-h\left(\frac{L_t}{L_s}\right)^2\right]
  &&\text{1st order}\nonumber\\
  \frac{T_c(\infty)-T_c(V)}{T_c(\infty)}&\sim\frac{1}{V^\frac{1}{2\nu}}&&\Rightarrow
  &&\beta_c(V)=\beta_c(\infty)\left[1-k\left(\frac{L_t}{L_s}\right)^\frac{1}{\nu}\right]
  &&\text{2nd order}
\label{phase:infinitevolume}
\end{align}
where $h,k$ are constants and we use  $T=1/(aL_t)=\beta g^2/(2NL_t)$.
In 2 spatial dimensions, the maximum of the susceptibility peak 
$\chi_{\text{max}}(V)$ depends on the spatial volume $V$ as
\begin{align}
&&\chi_{\text{max}}(V)&=c_0V+c_1 &&\text{1st order}&&\nonumber\\
&&\chi_{\text{max}}(V)&=c_0V^{\frac{\gamma}{2\nu}}+c_1 &&\text{2nd order}&&
\end{align}
for constants $c_0$ and $c_1$.
Hence, finite size scaling shows us how $\beta_c(V)$ and $\chi_{\text{max}}(V)$
vary with the spatial volume $V$, and how to extrapolate $\beta_c(V)$ to
the infinite volume limit.

\subsection{Useful order parameters}

An order parameter for a phase transition is a quantity that distinguishes between 
the different phases and exhibits a non-analyticity at the transition, and it
is this behaviour that allows us to determine if and where the deconfinement
phase transition occurs. 
As remarked above, phase transitions correspond to non-analyticities in the 
derivatives of the partition function $Z$ with respect to $\beta$. So consider 
the first two derivatives for our lattice action
\begin{align}
  \left(\frac{1}{N_p}\frac{\partial}{\partial\beta}\right)\text{ln}Z
  &\sim \braket{\overline{U}_p} \nonumber\\
  \left(\frac{1}{N_p}\frac{\partial}{\partial\beta}\right)\braket{\overline{U}_p}
  &=\braket{\overline{U}_p^2}-\braket{\overline{U}_p}^2
  \equiv\chi_{\overline{U}_p}/V
\label{phase:freeenergyderivatives}
\end{align}
where $N_p$ is the number of plaquettes, 
$\overline{U}_p=\tfrac{1}{N_p}\sum_p\left(\tfrac{1}{N}\text{tr}(U_p)\right)$ is
the plaquette averaged over the lattice volume, and 
$\chi_O=V\left(\braket{O^2}-\braket{O}^2\right)$ is the susceptibility of the operator $O$.
In the case of a first order transition we expect $\braket{\overline{U}_p}$ to
exhibit a finite discontinuity at $T=T_c$, and $\chi_{\overline{U}_p}$ to be
a $\delta$-function when $V\to\infty$. For a second order transition $\braket{\overline{U}_p}$
will be continuous, but  will have a divergent first derivative at $T_c$
when $V\to\infty$, so that $\chi_{\overline{U}_p}$ will display a divergence as
described above. Thus ${\overline{U}_p}$ appears
to be the obvious order parameter for locating the phase transition. 

\begin{figure}[h]
	\centering
  	\includegraphics[width=0.9\textwidth]{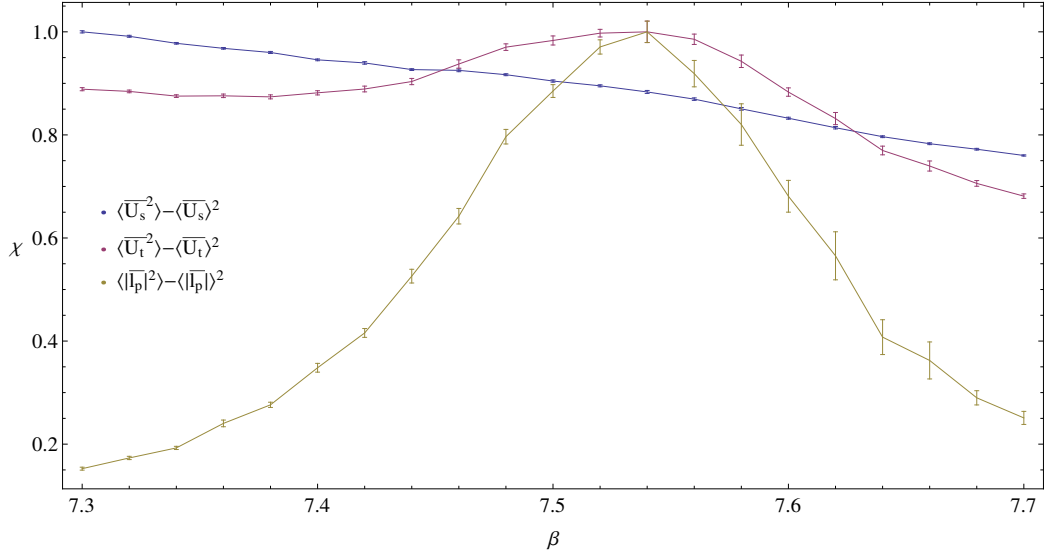} 
	\caption{Susceptibility plots (renormalised) for the spatial plaquette
          $\overline{U}_s$, the temporal plaquette $\overline{U}_t$, and the
          modulus of the averaged Polyakov loop $\left| \overline{l_P}\right|$ for $SO(4)$
          on a $32^2 3$ lattice.}
	\label{phase:fig:orderparameter}
\end{figure}

Unfortunately, our calculations indicate that the plaquette susceptibility has a
weakly varying signal over the phase transition -- too weak in fact to be useful on
the lattice volumes that we are able to contemplate using. To show what happens it
is convenient to partition the plaquettes into those that are only spatial
$\overline{U}_s$ and those that have links in a temporal direction $\overline{U}_t$.
Figure~\ref{phase:fig:orderparameter} shows the spatial plaquette susceptibility
$\chi_{\overline{U}_s}$ and the temporal plaquette susceptibility $\chi_{\overline{U}_t}$
in the region of the phase transition for an $SO(4)$ $32^23$ volume (renormalised
for purposes of comparison).
We need a clear peak in the susceptibility to identify the location of the phase
transition but we see instead that $\chi_{\overline{U}_s}$ has no obvious peak
structure while $\chi_{\overline{U}_t}$ has only a very weak peak structure.
For other $SO(N)$ groups, we also typically find that $\chi_{\overline{U}_{s,t}}$ have
no useful peak structures on the volumes we use. Of course when 
$L_s\to\infty$ the peaks should eventually appear and grow, but it does mean that
for our purposes the plaquette susceptibility is not a useful order parameter.

An alternative order parameter is provided by the temporal Polyakov loop $l_P$.
As described earlier, its expectation value has a direct relation to
the free energy of an isolated charge, and it is therefore a natural
order parameter for the deconfining transition.
We shall shortly see that the Polyakov loop operator $\overline{l_P}$ has a
much clearer signal in the region of the phase transition, compared to the
plaquette operators.
Around the transition it tunnels between confined and deconfined phases so that 
$\overline{l_P}$ takes discrete values with very small fluctuations around these.
There is however a problem at finite $V$. If there is a non-trivial centre
symmetry then tunnelling between the corresponding deconfined phases will
cause ${\overline{l_P}}$ to average to zero for $T>T_c$. This is not an issue 
for $SO(2N+1)$ gauge theories since these have a trivial center symmetry,
but it is a problem for $SO(2N)$ with its $\mathbb{Z}_2$ center symmetry. The same
problem arises, of course, for $SU(N)$ gauge theories. The standard
(if theoretically ugly) fix is to take the absolute value of the Polyakov loop 
after averaging it over the spatial volume 
\begin{align}
\left| \overline{l_P}\right|=\left|\frac{1}{L_s^2}\sum_\mathbf{x} l_P(\mathbf{x})\right|
\end{align}
and to use $\left| \overline{l_P}\right|$ as an order parameter and to
construct an associated susceptibility from that,
\begin{align}
  \frac{\chi_{\left| \overline{l_P}\right|}}{L_s^2L_t}
  =\Braket{\left| \overline{l_P}\right|^2}-\Braket{\left| \overline{l_P}\right|}^2.
\label{phase:susceptibility}
\end{align}
This has the disadvantage that $\Braket{\left| \overline{l_P}\right|}\neq 0$
in the confined phase as well as in the deconfined phase, but the values are
very different and it has a very good signal in the region of the phase transition.
We return to our $SO(4)$ $32^23$ lattice in Figure~\ref{phase:fig:orderparameter},
and plot the Polyakov loop susceptibility.
We see that $\chi_{\left| \overline{l_P}\right|}$ has a much clearer peak structure
than the plaquette susceptibilities shown in the same figure.

\begin{figure}
	\centering
  	\includegraphics[width=\textwidth]{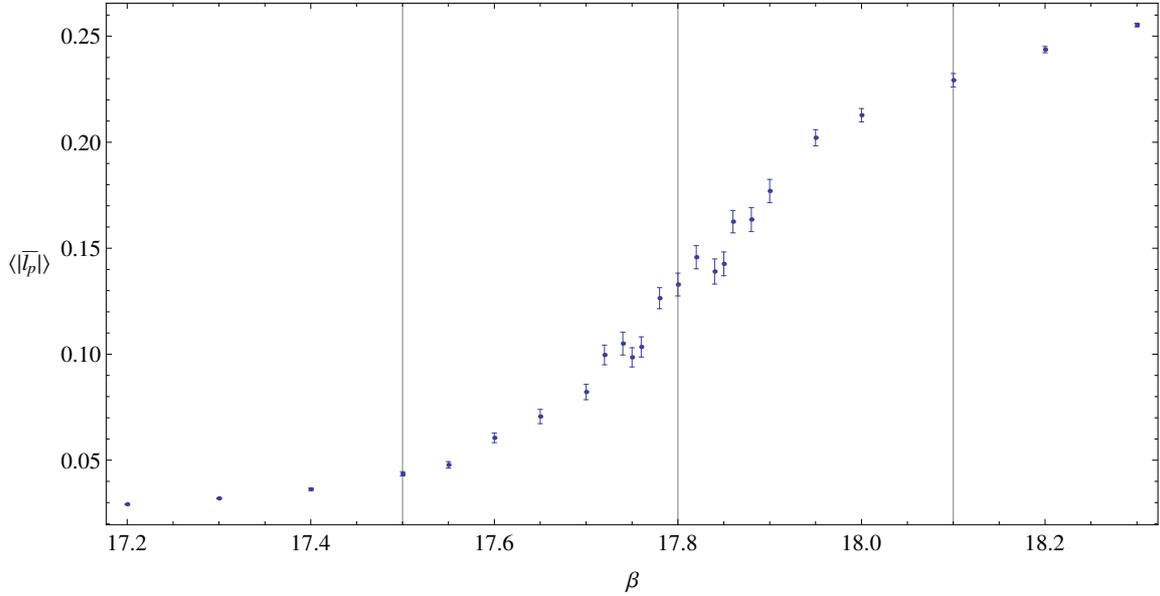} 
	\caption{The Polyakov loop $\Braket{\left| \overline{l_P}\right|}$ for
          $SO(6)$ on a $20^2 3$ lattice. The vertical lines, spanning the
          transition, correspond to
          $\beta=(\beta_{-},\beta_0,\beta_{+})=(17.5,17.8,18.1)$.}
	\label{phase:fig:transitionexamplejump}
\end{figure}

So, in a plot of $\Braket{\left| \overline{l_P}\right|}$ against
$\beta$ in the neighbourhood of $\beta_c$, we would expect to see the value
of $\Braket{\left| \overline{l_P}\right|}$ increase from near-zero to some
non-zero value over a narrow range of $\beta$. For a first order transition
this range shrinks to zero as the volume increases, becoming a discontinuity
at $V=\infty$, while for a second order transition this range remains finite
and there is no discontinuity, but the slope at $\beta_c$ tends to $\infty$.
We show an example, obtained on a $20^2 3$ lattice in $SO(6)$,
in   Figure~\ref{phase:fig:transitionexamplejump}. We expect to see a 
corresponding peak in $\chi_{\left| \overline{l_P}\right|}$ at $\beta_c$,
as in  Figure~\ref{phase:fig:orderparameter}. For a first order 
transition, we expect the susceptibility $\chi_{\left| \overline{l_P}\right|}$ 
to approach a delta function singularity as $V\to\infty$. 
For a second order phase transition, we expect that the
susceptibility $\chi_{\left| \overline{l_P}\right|}$ has a peak over
a finite range of $\beta$ around $\beta_c$, with a cusp-like divergence at
$\beta_c$.

\begin{figure}[htb]
\begin	{center}
\input	{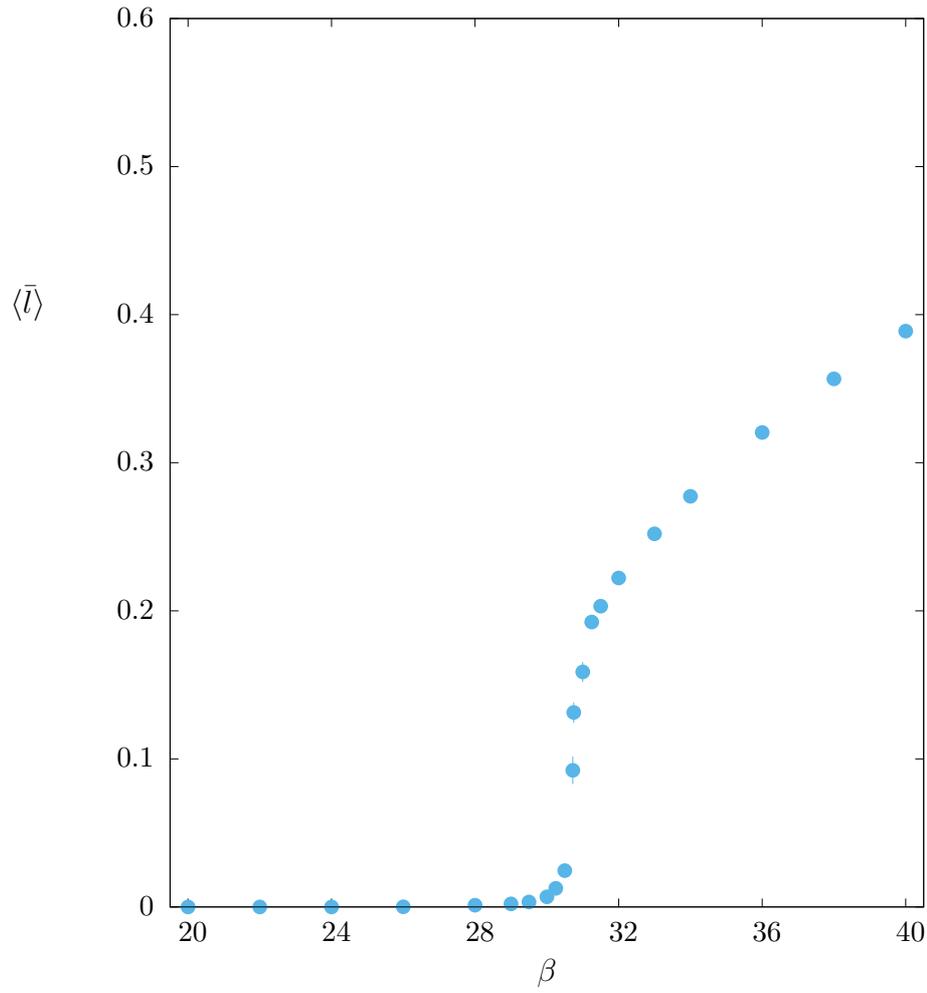}
\end	{center}
\caption{Average value of Polyakov loop on a $48^2 4$ lattice in $SO(7)$ versus
$\beta$. }
\label{polyso7_t4l48}
\end{figure}

For odd $N$ there is no $Z_2$ symmetry to be spontaneously broken, so we
can use our cleaner original variable, $\Braket{\overline{l_P}}$, to characterise 
the transition. In Fig.\ref{polyso7_t4l48} we plot this quantity against $\beta$ 
for a $48^24$ lattice in $SO(7)$ with a sharp transition visible near the middle 
of the range. (Since the lattice spacing varies roughly as
$1/\beta$, the range $\beta \in [20,40]$ corresponds roughly to the range
$T/T_c\in [0.66,1.5]$.) Despite the lack of a centre symmetry, we find
that for $\beta \leq 26.0$ our values are all consistent with 
$\Braket{\overline{l_P}}$ being zero within errors, with values
$\sim \pm 10^{-5}$. This behaviour motivates describing the transition
as being `deconfining' even if the low-$T$ vacuum eventually turns out
not to be exactly confining.

\subsection{Tunnelling}

We can represent the values of $\overline{l_P}$ obtained from the sequence of 
field configurations generated at a given $\beta$ in a Monte Carlo run 
as either a histogram over the entire run, or as a history plot along the run. 
For $\beta<\beta_c$, we expect the theory to be confining so that 
$\Braket{\overline{l_P}}\approx0$. On the histogram, we would 
expect that the values of $\overline{l_P}$ form a narrow peak around zero 
while, on the history plot, we would expect the values to fluctuate around zero. 
For $\beta>\beta_c$, the system would be in a deconfined phase so that 
$\Braket{\overline{l_P}}\neq0$, and we would expect to see deconfined peaks at 
non-zero values on the histogram.
For $SO(2N)$ gauge theories, we would expect to see two deconfined peaks at non-zero 
values, reflecting the spontaneous breaking of the $\mathbb{Z}_2$ center symmetry, while, for 
$SO(2N+1)$ gauge theories, where the center symmetry is trivial, we would only expect 
one deconfined peak at a non-zero value. 

\begin{figure}
	\centering
  	\includegraphics[width=\textwidth]{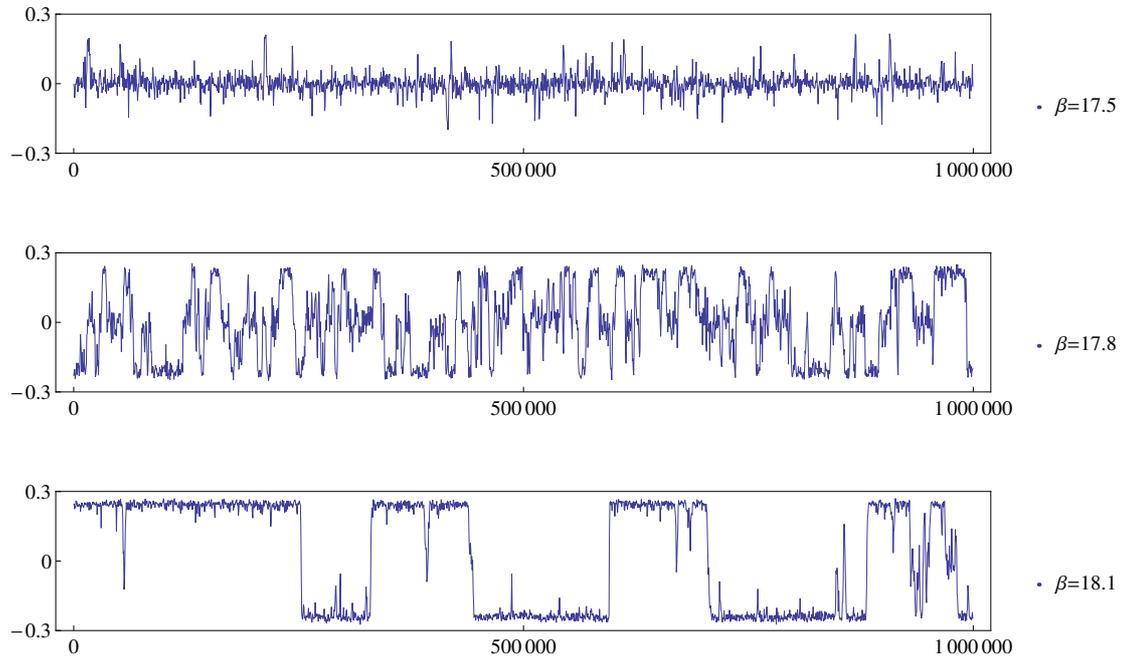} 
	\caption{Polyakov loop $\overline{l_P}$ history plots at
          $\beta=\beta_{-},\beta_0,\beta_{+}$ (see
          Figure~\ref{phase:fig:transitionexamplejump}) for $SO(6)$ on a
          $20^2 3$ lattice from a run of $10^6$ configurations.}
	\label{phase:fig:transitionexamplehistory}
\end{figure}

\begin{figure}[htb]
\begin	{center}
\input	{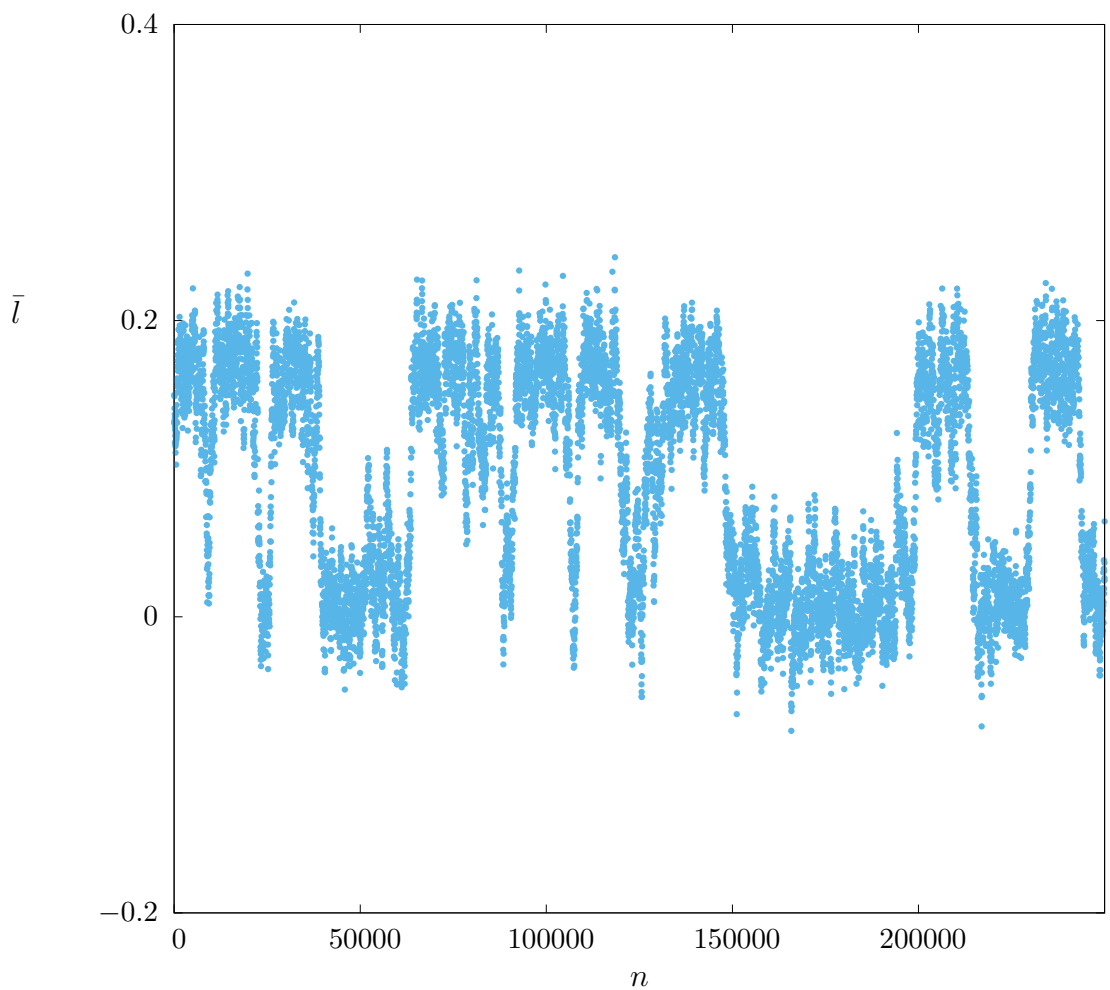}
\end	{center}
\caption{The value of the averaged Polyakov loop, taken every 25 sweeps, on a 
$48^2 4$ lattice in $SO(7)$ versus the number of sweeps $n$. }
\label{runplso7_t4l48}
\end{figure}

\begin{figure}[htb]
\begin	{center}
\input	{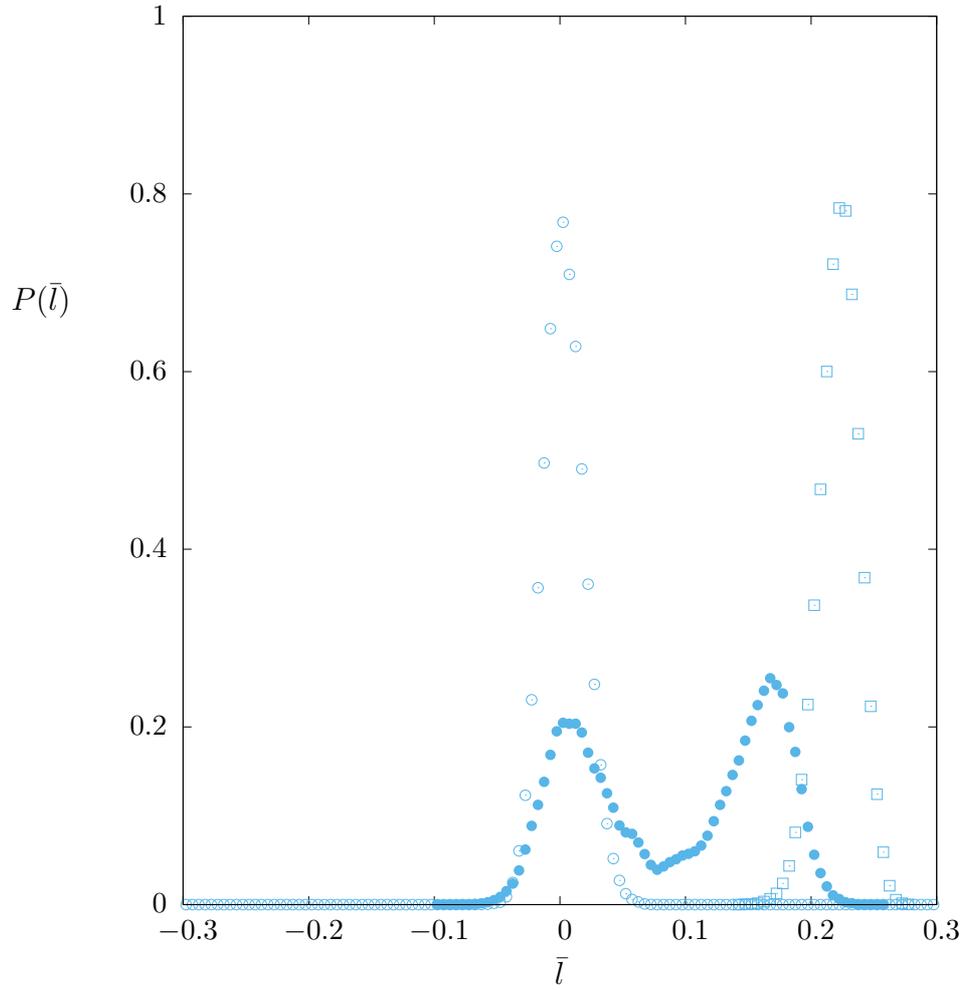}
\end	{center}
\caption{Probability (unnormalised) of the Polyakov loop averaged over the spatial volume of a $48^24$ lattice in $SO(7)$ for $T\simeq 0.96 T_c$, $\circ$, $T\simeq T_c$, $\bullet$, and $T\simeq 1.04 T_c$, $\Box$.}
\label{histso7_t4l48}
\end{figure}

For a first order transition we would expect that as we increase $\beta$ towards 
$\beta\approx\beta_c$, and beyond, we should see deconfined peaks appear at non-zero 
values while the confined peak at zero decreases. And in a history plot we would 
see jumps that reflect tunnelling between the confined and deconfined phases. 
Beyond $\beta\approx\beta_c$ any tunnelling should be only between 
the two deconfined phases for even $N$, and no tunnelling for odd $N$. The 
behaviour for even $N$ is illustrated for $SO(6)$ on a $20^2 3$ lattice 
in the histograms in Fig.\ref{phase:fig:transitionexamplehistogram} and 
the history plots in Fig.\ref{phase:fig:transitionexamplehistory}.
For odd $N$ we illustrate the expected behaviour in  $SO(7)$ on a $48^2 4$ lattice 
in the history plot in Fig.\ref{runplso7_t4l48} and the histograms
in Fig.\ref{histso7_t4l48}. The coexistence of both confining and deconfining peaks 
at a given $\beta$ establishes that we have a first order transition in
both $SO(6)$ and $SO(7)$.

\begin{figure}
	\centering
  	\includegraphics[width=\textwidth]{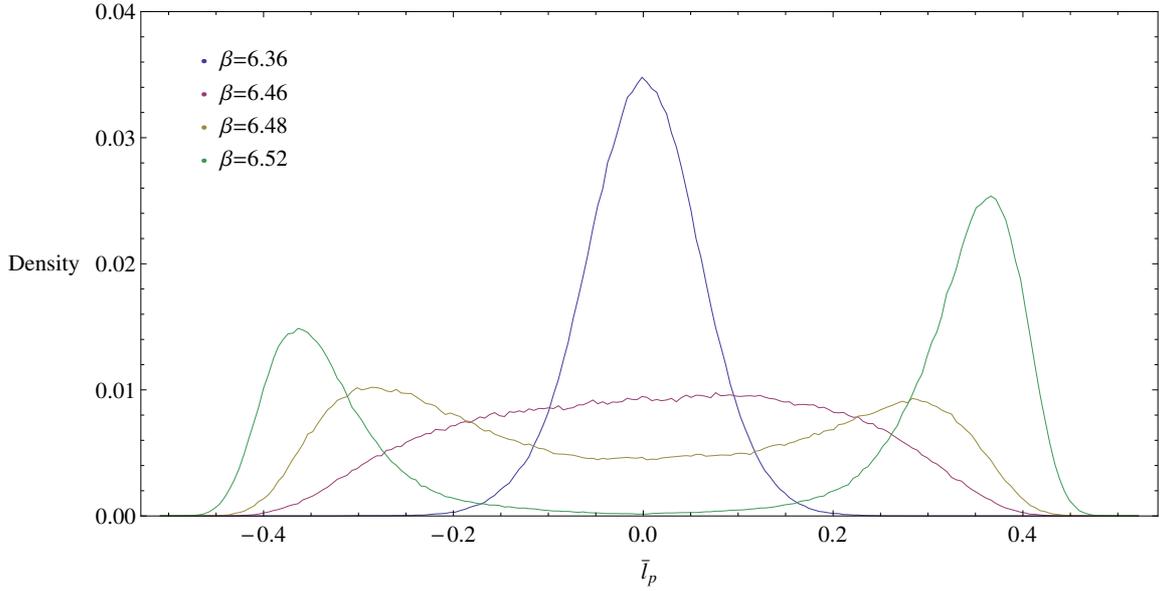} 
	\caption{Polyakov loop $ \overline{l_P}$ histograms for $SO(4)$
          on a $28^2 2$ lattice.}
	\label{phase:fig:so4histogram}
\end{figure}

For a second order transition, there is no phase coexistence. 
As we increase $\beta$, we would expect the confined peak around zero to spread 
out and, once it disappears, the deconfined peaks emerge at $\beta=\beta_c$. 
On the history plot, we would expect to see significant fluctuations around zero for 
$\beta<\beta_c$ before the onset of tunnelling between the deconfined phases 
for $\beta>\beta_c$. This is illustrated for the case of a $28^2 2$ lattice
in $SO(4)$ in Figure~\ref{phase:fig:so4histogram}.

Hence, we can use both the histograms and history plots of $\overline{l_P}$ 
to distinguish between first and second order transitions.

\subsection{Identifying $\beta_c$}

\begin{figure}[h]
	\centering
  	\includegraphics[width=0.9\textwidth]{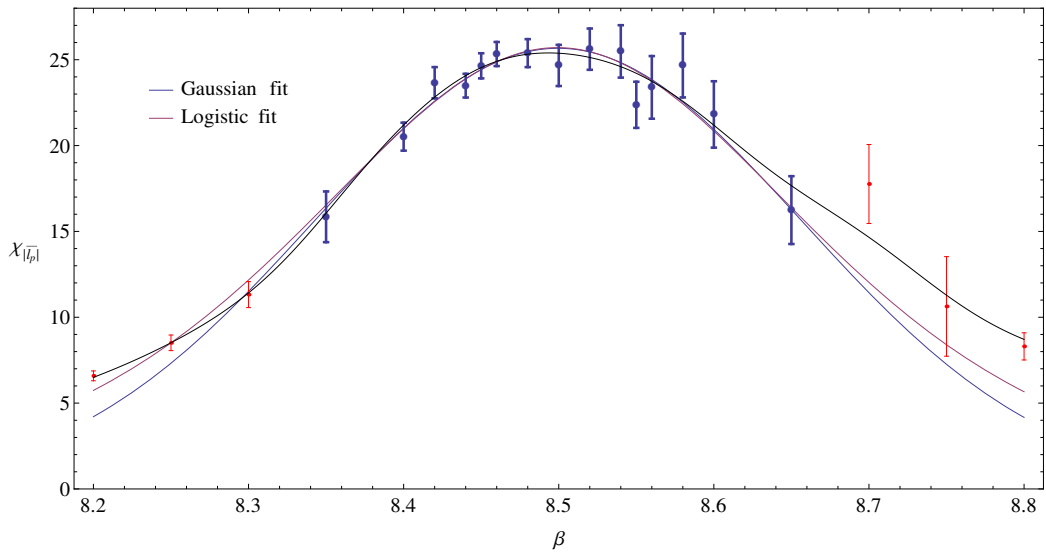} 
	\caption{Fitting the Polyakov loop susceptibility in $SO(4)$ on a $40^24$ lattice.
          The points represent our calculations while the continuous black line
          represents reweighted values. Other lines are curve fits, using the
          data from the dark points rather than the light points to reduce
          the $\bar{\chi}^2_\text{dof}$ of the fits.}
	\label{phase:fig:curvefitting}
\end{figure}

To calculate  $\beta_c$ on a given volume $V$ we need to locate the maximum of
the susceptibility. We do so by first performing separate runs at different
$\beta$ values, and then doing more runs at values of $\beta$ near the peak.
We use the standard density of states reweighting method 
\cite{Ferrenberg:1988nr,Huang:1991lr}
to construct a smooth interpolating function through the measured values, 
whose maximum provides our estimate of $\beta_vc$ on the given volume $V$.
For some very large spatial volumes, the values that arise in the reweighting 
algorithm exceed the machine precision. In principle this obstacle should be
surmountable by some judicious alteration of the algorithm, but 
in these cases we choose instead to use curve fitting to find $\beta_c$,
based on a logistic function for the Polyakov loop, 
which in practice turns out to have a comparable performance to that
of our reweighting algorithm, as we see from 
Fig.\ref{phase:fig:curvefitting} and Table~\ref{phase:tab:so4curvefitcomparison}.

\section{$SO(N)$ lattice calculations in $D=2+1$}
\label{sec:lattice}

We generate sequences of lattice field configurations using an $SO(N)$ adaptation 
of the $SU(N)$ Cabbibo-Marinari heat bath algorithm 
\cite{Cabibbo1982387}, 
which we describe in our companion paper on the $SO(N)$ spectrum,
\cite{Lau:aa}.
We use the plaquette action in eqn(\ref{eqn:plaqaction}).

We express the deconfining temperature in physical units by calculating suitable 
mass scales $\mu$ of the gauge theories at $T=0$ and then taking ratios 
$aT_c/a\mu = T_c/\mu$, which we can then extrapolate to the continuum limit in a 
standard way. Three such quantities are the string tension, coupling, and lightest 
scalar glueball mass (the mass gap). We now briefly describe how we calculate these on 
the lattice. We provide fuller details in 
\cite{Lau:aa}.

\subsection{String tensions}

To obtain the string tension, we calculate the energy $E(l)$ of the lightest 
flux tube that winds around the spatial torus of size $l$ on a lattice
that corresponds to $T\sim 0$. To do this, we use 
correlators of zero-momentum sums of Polyakov loop operators, that have been 
`blocked' to obtain a very good overlap onto the ground state
\cite{Teper1987345}
supplemented by a standard variational calculation
\cite{Teper:1998rw}. 
We expect that $E(l)\rightarrow \sigma l$ for $l$ large
\cite{Lau:aa}. 
For finite $l$, we expect $E(l)$ to be well-approximated by
\cite{Athenodorou:2011hl,Aharony,Dubovsky} 
\begin{align}
E(l) = \sigma l \left(1-\frac{\pi}{3\sigma l^2}\right)^{\frac{1}{2}} .
\end{align}
By evaluating the string tension at $\beta_c$, we can then express the deconfining 
temperature in the dimensionless ratio $T_c/\surd\sigma$.

\subsection{Couplings}

In $D=2+1$ the coupling $g^2$ provides a mass scale for the theory.
In the continuum limit 
\begin{align}
\lim_{\beta \to \infty} \frac{\beta}{2N^2}=\frac{1}{ag^2N}
\end{align}
where $g^2N$ is the 't Hooft coupling which one keeps constant as $N$ increases
in order to have a smooth large-$N$ limit. At finite lattice spacing the coupling
is scheme dependent, and in that sense not a physical quantity, but different
choices of coupling differ at $O(ag^2)$ and so converge to the same continuum limit.
It makes sense to try and choose a coupling scheme within which that convergence
is rapid. Previous calculations in $D=2+1$ $SU(N)$
\cite{Teper:1998rw}
have found it useful to employ the mean field improved coupling
 \cite{Lepage:1993dp}
\begin{align}
\beta_I =\beta \Braket{\frac{1}{N}\text{tr}(U_p)} .
\end{align}
We will choose to use this improved coupling to calculate the continuum
value of $T_c/g^2N$.

\subsection{Scalar glueball masses}

$SO(N)$ gauge theories have a glueball mass spectrum similar to that in $SU(N)$ gauge 
theories, except that all glueballs have charge conjugation $C=+$.
The lightest glueball has spin $J=0$ and parity $P=+$ and it is the glueball 
mass that we can calculate most accurately.
We evaluate the continuum glueball masses $M_{0^+}/\surd\sigma$ in 
\cite{Athenodorou:2015aa,Lau:aa} 
and use these values as another way of expressing the deconfining temperature in 
physical units $T_c/M_{0^+}=T_c/\surd\sigma \times \surd\sigma/M_{0^+}$.

\section{Results: infinite volume limits}
\label{sec:infinitevolume}

\subsection{Methodology}

\begin{figure}
	\centering
  	\includegraphics[width=\textwidth]{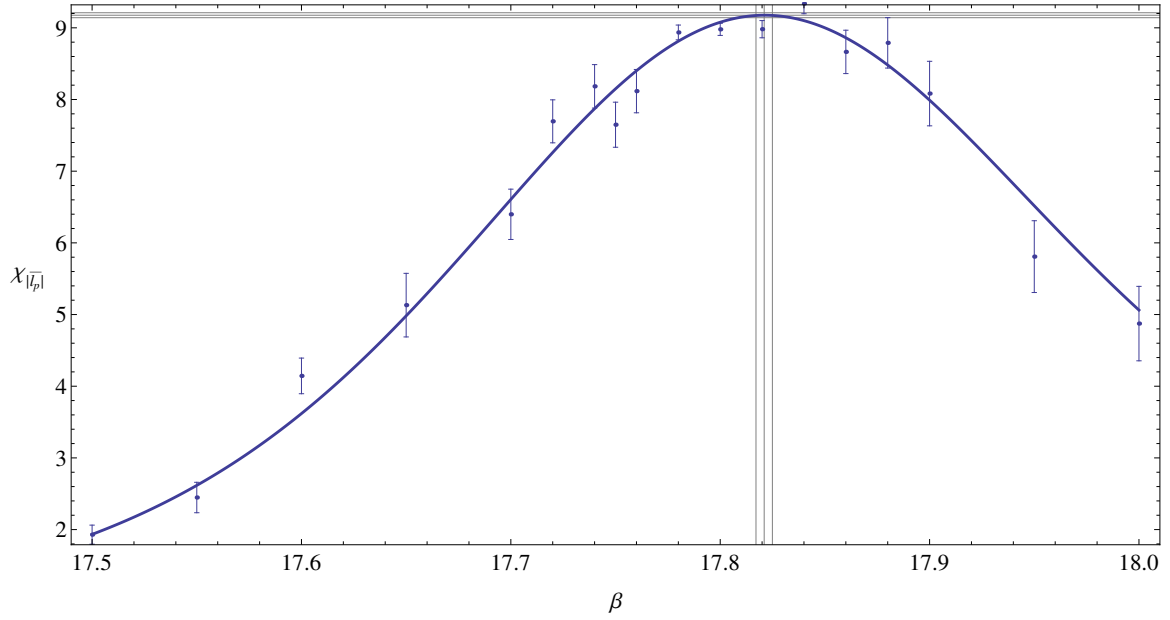} 
	\caption{The reweighted susceptibility compared to our data, for a
          $20^2 3$ lattice in $SO(6)$. The vertical and horizontal lines correspond
          to the values at the maximum of the susceptibility with its error:
          $\beta_c=17.821(4)$ and $\chi_{\left| \overline{l_P}\right|}(\beta_c)=9.18(3)$}
	\label{phase:fig:transitionexamplereweighting}
\end{figure}

We need to calculate $\beta_c(V\rightarrow\infty)$ on our $L_s^2 L_t$ lattices, 
to obtain the lattice deconfining temperature $T_c = 1/a(\beta_c(V=\infty))L_t$.
Using $\left| \overline{l_P}\right|$ as our order parameter, for a given finite spatial
volume $V$, we calculate $\beta_c(V)$ by calculating the susceptibility
$\chi_{\left| \overline{l_P}\right|}$ for a range of $\beta$ values, reweighting the 
data from those $\beta$ values where we observe there to be tunnelling between 
the confined and deconfined phases, and then locating the maximum.
If the lattice volume is too large for our reweighting algorithm, we follow the curve
fitting procedure mentioned above.
Then $\beta_c(V)$ is the $\beta$ value that corresponds to a maximum in
$\chi_{\left| \overline{l_P}\right|}$. This is illustrated in  
Fig.~\ref{phase:fig:transitionexamplereweighting} on a $20^2.3$ lattice in
$SO(6)$ where we see that the reweighted curve agrees well with our original data 
and that the estimates for $\beta_c$ and $\chi_{\left| \overline{l_P}\right|}(\beta_c)$ 
have very small errors. 

\begin{figure}
	\centering
  	\includegraphics[width=\textwidth]{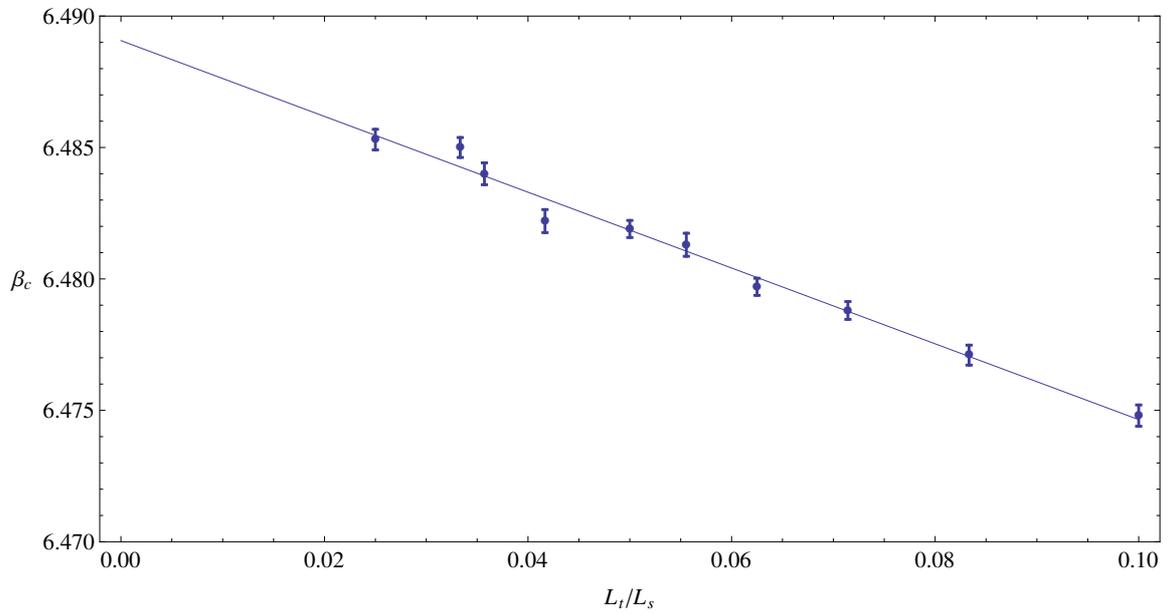} 
	\caption{The infinite volume extrapolation for $SO(4)$ with $L_t=2$.}
	\label{phase:fig:so4finitevolume}
\end{figure}

\begin{figure}
	\centering
  	\includegraphics[width=\textwidth]{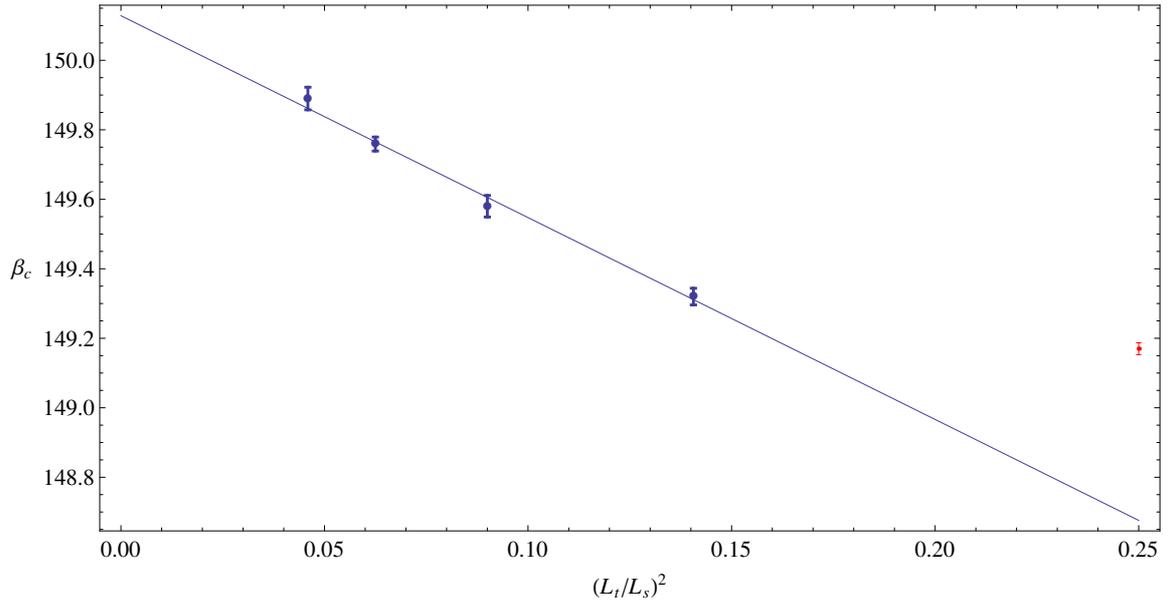} 
	\caption{The infinite volume extrapolation (using dark points only) for $SO(16)$
          with $L_t=3$ and with a range of $L_s$ values.}
	\label{phase:fig:so16finitevolume}
\end{figure}

Repeating this calculation for a range of $V$ we can extrapolate to $V=\infty$ using 
the finite size scaling formulae in eqn(\ref{phase:infinitevolume}).
In Figure~\ref{phase:fig:so4finitevolume} we display such an infinite volume
extrapolation for a second order transition in $SO(4)$ with $L_t=2$,
and in Figure~\ref{phase:fig:so16finitevolume} for a first order transition,
in $SO(16)$ with $L_t=3$.
In both cases we see that the extrapolation is precise and well-defined.
As will be apparent when we list the results of our extrapolations,
this is mostly the case, albeit with a significant number of
exceptions where the fits are statistically poor .

Since the tunnelling in a first order transition is important to both
identifying and locating the transition, it is useful to consider 
how this tunnelling varies with $V$ and $N$. Using a standard 
argument, the tunnelling must proceed through an intermediate 
configuration where the two phases are separated by two spatial 
domain walls of length $l_s=aL_s$, with a probability of
\begin{align}
  P_W(T)\propto\text{exp}\left(-\frac{2\sigma_W l_s}{T}\right)
  =\text{exp}\left(-2a^2\sigma_W L_sL_t\right)
\end{align}
relative to the probability of a single phase at the same temperature.
Here $\sigma_{W}$ is the surface tension per unit length of the domain wall. 
(All this assumes that our Monte Carlo is a local process. If we have global
updates, which are trivial to construct between the two deconfined phases, then
this discussion will need changing.) Now just as in $SU(N)$ we expect 
the surface tension to grow with $N$ as $\sigma_W\propto N^2$
\cite{Lucini:2004fk}. 
Hence, the probability of the domain walls and the probability of tunnelling
decreases exponentially as either the volume $V$ or as $N$ increase. 
Thus transitions between the two states are increasingly rare at large $V$,
especially at large $N$, and this provides an effective upper bound on the
volumes we can consider at a given $N$. In addition to this, critical
slowing down will also suppress the frequency of tunnelling as $a(\beta_c)$
decreases. 

Since the accuracy of our calculation of $\beta_c(V)$ depends 
primarily on the number of tunnelling fluctuations, rather than the
fluctuations within a given phase, we should, ideally, use errors 
in our reweighting procedure derived solely from the number of 
tunnellings. Since this is not straightforward to do, we instead
used only data points from runs that clearly have tunnellings,
but then used `naive' errors, albeit based on large bin-sizes
each of which would usually contain some tunnellings. 
While we believe this `fix' is usually reliable, it
nonetheless  leaves a systematic error in our
calculations which we only partially control, and this may 
be the reason for the very poor goodness of fit of a few of
our $V\to\infty$ extrapolations.

\subsection{$SO(4)$ and $SO(5)$}

\begin{figure}
	\centering
  	\includegraphics[width=\textwidth]{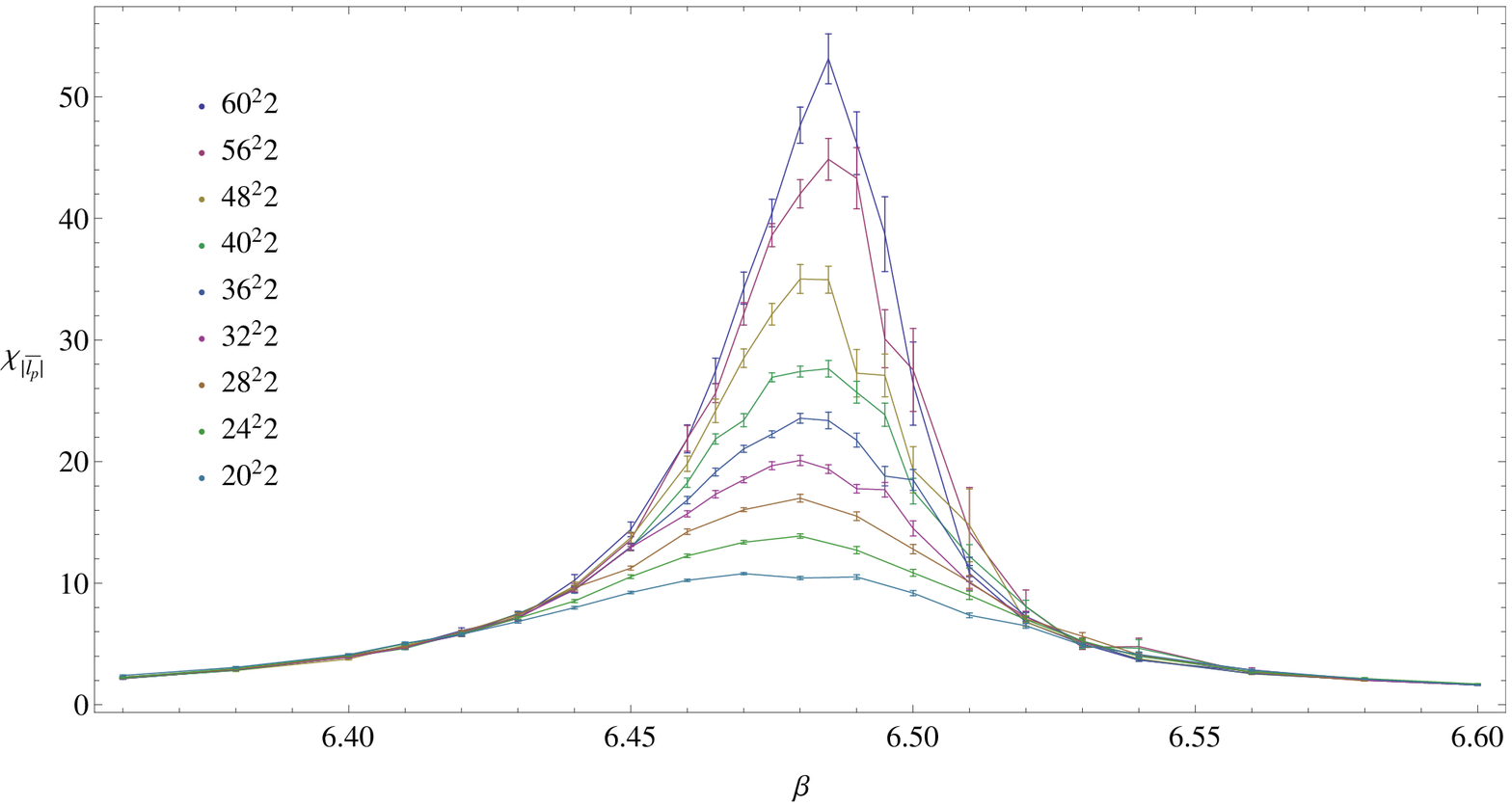} 
	\caption{Susceptibility volume dependence for $SO(4)$ and $L_t=2$.}
	\label{phase:fig:so4susceptibility}
\end{figure}

The $SO(4)$ and $SO(5)$ deconfining phase transitions are second order. 
We can see this from the ${ \overline{l_P}}$ histograms, such as
Figure~\ref{phase:fig:so4histogram}, which show a continuous transition
from confined to deconfined phases as we increase $\beta$. 
We can also see this in susceptibility plots for different spatial volumes
at fixed $L_t$, such as Figure~\ref{phase:fig:so4susceptibility}, which
show that, as the spatial volume increases, the susceptibility peak height
increases, and the large volume susceptibility provides an envelope for 
the ones at smaller $V$.

For $SO(4)$, we can use reweighting for $2\leq L_t\leq4$ to calculate $\beta_c$. 
For $L_t=5$, the susceptibility peak is at $\beta\in[9.0,10.0]$.
This is in the region of the `bulk' transition which separates weak 
and strong coupling and which we will discuss later, and which affects
the data so greatly that reweighting does not work. 
For $L_t\geq 6$, the spatial volumes become so large that we cannot reweight
the data using our standard algorithm and so we curve fit instead. 
For smaller $L_t$, the values lie on a smooth curve with small errors and
the reweighted values fit well with the original data. 
At larger $L_t$, the data is more scattered than at smaller $L_t$, although
we can still estimate $\beta_c$ with usefully small errors. 
We present the $SO(4)$ values of $\beta_c(V)$ for volumes $V$ with
$L_t=2,3,4,6,7,8,10,12$ in Tables~\ref{phase:tab:so4susceptibilitydatapart1}
and \ref{phase:tab:so4susceptibilitydatapart2}.
 
To extrapolate $\beta_c(V\rightarrow\infty)$ to the infinite volume limit
using eqn\eqref{phase:infinitevolume}, we need a value for the critical
exponent $\nu$.
We recall that the  Svetitsky-Yaffe conjecture
\cite{Svetitsky1982423}
puts the deconfining phase
transition in the same universality class as the order/disorder transition
of the spin system which is in the same spatial dimensions and which is
invariant under the group that corresponds to the centre of the 
gauge group. For $SO(2N)$ gauge groups, which have a $\mathbb{Z}_2$ centre
symmetry, this puts the deconfining phase transition in the same universality
class as the $D=2$ Ising model.
In the case of $SO(4)\sim SU(2)\times SU(2)$ we would expect the deconfining
phase transition to be in the universality class of two decoupled $D=2$ Ising models.
In the case of $SO(5)$, we know that $Sp(2)$ forms the vector representation of
$SO(5)$, which also has a $\mathbb{Z}_2$ centre symmetry so we would expect that 
its deconfining phase transition should also be in the universality class of
the $D=2$ Ising model
\cite{Pepe:2004}.
Since the order/disorder transition for the $D=2$ Ising model has critical
exponents
\begin{align}
\gamma=1.75 \quad ; \quad \nu=1
\end{align}
we expect these to be the critical exponents of the $SO(4)$ and $SO(5)$
deconfining phase transitions.
One can try to support this choice by fitting $\nu$ to our actual data, but because
the variation of $\beta_c(V)$ is weak one needs a large lever arm in $V$, and
very accurate data, to get a useful result. With our data the only useful
fit for $\nu$ is to the $SO(4)$ $L_t=2$ data from which we obtain the estimate
$\nu=0.88(19)$, which provides some support for the universality based
value, which we shall employ from now on. 

We list the resulting $SO(4)$ $\beta_c(V=\infty)$ values in
Table~\ref{phase:tab:so4infinitevolume} showing in each case the goodness 
of fit as measured by the value of $\bar{\chi}^2_\text{dof}$ (chi-squared divided 
by the number of degrees of freedom). We see that the extrapolated values
have small errors and most of the $\bar{\chi}^2_\text{dof}$ values are
reasonable. (One $\bar{\chi}^2_\text{dof}$ value is very large, and this is due
to a scatter among values with very small errors, which cannot be remedied
by dropping values at the smallest $V$.)

For $SO(5)$, we can use reweighting for $2\leq L_t\leq6$ and curve fitting
for $L_t\geq7$ to calculate $\beta_c$. 
Since the centre symmetry is trivial we cannot use that to argue that
$\Braket{\overline{l_P}}\approx 0$ in the low $T$ confined phase. 
However, our calculations show that this is indeed the case.
We list the $\beta_c(V)$ values for $SO(5)$ with $L_t=2,3,4,5,6,7,8,10$
in Table~\ref{phase:tab:so5susceptibility} and for the infinite volume limits
in Table~\ref{phase:tab:so5infinitevolume}. We see that the extrapolated values
again have small errors and that the $\bar{\chi}^2_\text{dof}$ values
are mostly reasonable.

\subsection{$SO(6)$}

\begin{figure}
	\centering
  	\includegraphics[width=\textwidth]{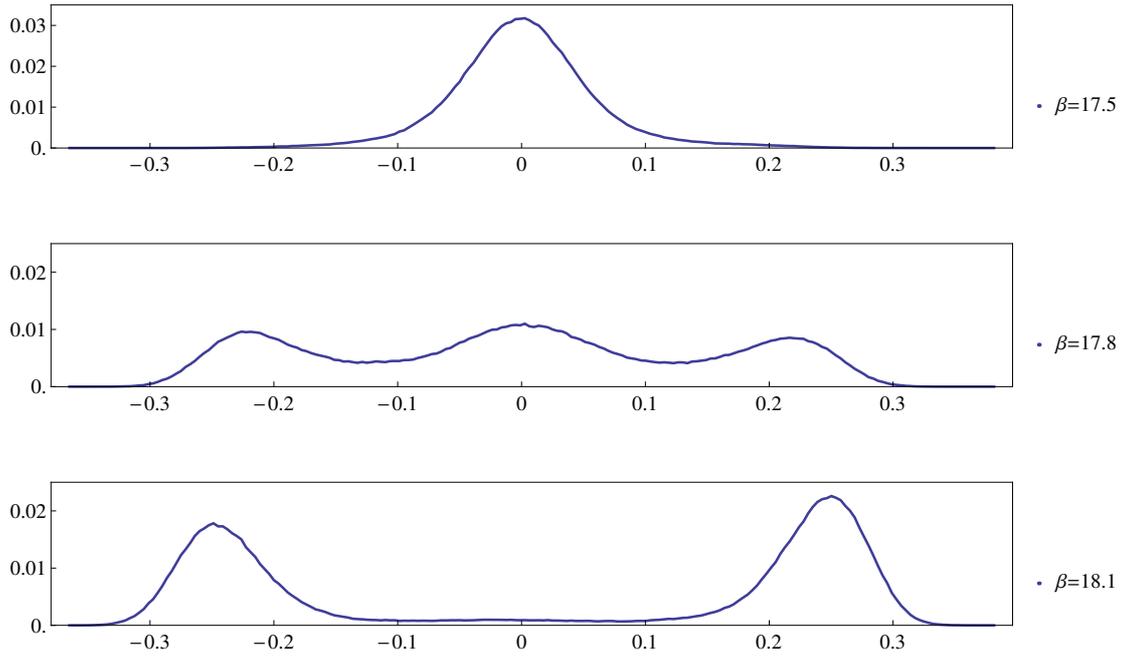} 
	\caption{Polyakov loop $\overline{l_P}$ histograms at
          $\beta=\beta_{-},\beta_0,\beta_{+}$ (see
          Figure~\ref{phase:fig:transitionexamplejump}) for $SO(6)$ on a $20^2 3$ lattice.}
	\label{phase:fig:transitionexamplehistogram}
\end{figure}

The $SO(6)$ deconfining phase transition is (weakly) first order: 
the coexisting phases are apparent, as in
Figure~\ref{phase:fig:transitionexamplehistogram},
but are less well defined than for $SO(N\geq 7)$. While susceptibility plots
indicate that the transition has features from both first and second order
transitions, the ${ \overline{l_P}}$ histograms (such as
Figure~\ref{phase:fig:transitionexamplehistogram}) show a clear first order 
phase coexistence. We extrapolate to the infinite volume limit using
eqn\eqref{phase:infinitevolume}.
We list the $\beta_c(V)$ values in Table~\ref{phase:tab:so6susceptibility}
and the infinite volume limits in Table~\ref{phase:tab:so6infinitevolume}.

\subsection{$SO(7)$, $SO(8)$, $SO(9)$, $SO(12)$, and $SO(16)$}

\begin{figure}
	\centering
  	\includegraphics[width=\textwidth]{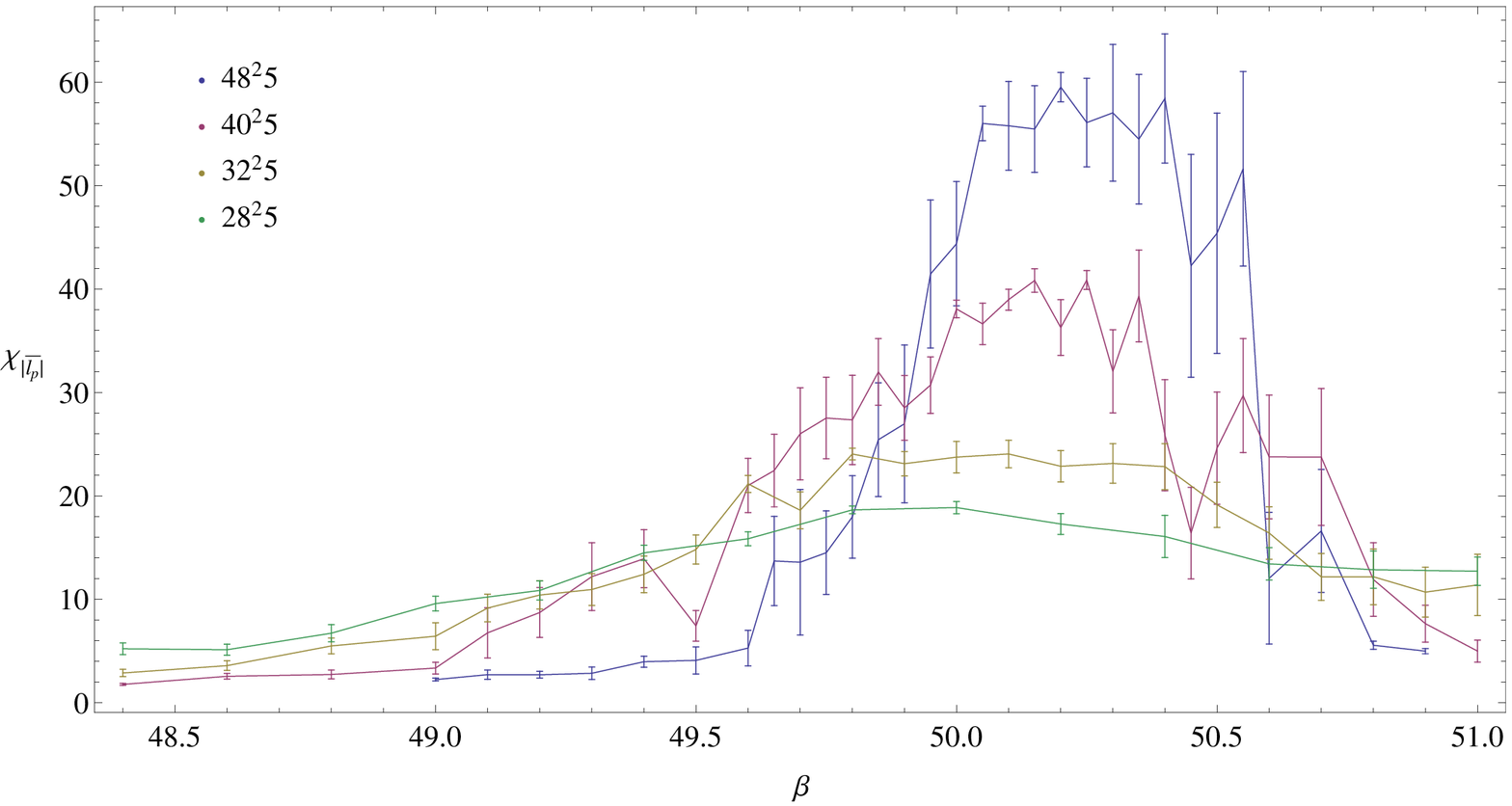} 
	\caption{Susceptibility volume dependence for $SO(8)$ and $L_t=5$.}
	\label{phase:fig:so8susceptibility}
\end{figure}

The $SO(N\geq 7)$ deconfining phase transitions are all first order, as is clear
from the phase coexistence in the $\Braket{ \overline{l_P}}$ histograms
and from the susceptibility plots (such as Figure~\ref{phase:fig:so8susceptibility})
which show the whole peak shrinking and its height growing as $V$ increases.
For $SO(7)$ and $SO(9)$ our calculations show that, just as for $SO(5)$, we have
$\Braket{ \overline{l_P}}\approx0$ despite the absence of a non-trivial
center symmetry. 

We list the $\beta_c(V)$ values in Tables~\ref{phase:tab:so7susceptibility}, 
\ref{phase:tab:so8susceptibility}, \ref{phase:tab:so9susceptibility}, 
\ref{phase:tab:so12susceptibility}, and \ref{phase:tab:so16susceptibility}
and the infinite volume limits, obtained using eqn\eqref{phase:infinitevolume},
in Tables~\ref{phase:tab:so7infinitevolume}, \ref{phase:tab:so8infinitevolume}, 
\ref{phase:tab:so9infinitevolume}, \ref{phase:tab:so12infinitevolume}, 
and \ref{phase:tab:so16infinitevolume}.

\section{Results: continuum limits}
\label{sec:continuum}

\subsection{Methodology}

To extrapolate $T_c$ to the continuum limit, i.e. $a\rightarrow 0$ or equivalently
$\beta \rightarrow \infty$, we express $T_c$ in units of some other energy scale 
$\mu$, calculated at the same value of $\beta$, and extrapolate the 
resulting  dimensionless ratio $\lim_{\beta \to \infty}T_c/\mu$. For the scale $\mu$ we
will use either the string tension, $\mu=\surd\sigma$,  calculated 
at $\beta_c$ and at $T\approx 0$, or the 't Hooft coupling, $\mu = g^2N$.  

Let us express the critical temperature in units of the string tension evaluated 
at the critical coupling $\beta_c$ on a lattice corresponding to  $T\simeq 0$, 
\begin{align}
\frac{T_c}{\surd\sigma}(a)=\frac{1}{a(\beta_c)\surd\sigma L_t} .
\end{align}
Once we have $T_c/\surd\sigma$ for each of our values of $L_t$, we take the continuum 
limit $a\rightarrow0$.
Since this is the ratio of two physical mass scales, we expect the leading correction 
to be $\mathcal{O}(a^2)$ \cite{Symanzik:1983uq},
\begin{align}
\frac{T_c}{\surd\sigma}(a)=\frac{T_c}{\surd\sigma}(a=0)+ca^2\sigma(a)+\cdots
\label{phase:stringcontinuumextrapolation}
\end{align}
for some constant $c$.

We can similarly express the critical temperature in terms of the 't Hooft coupling,
\begin{align}
\frac{T_c}{g^2N} \equiv \frac{aT_c}{ag^2N} = \frac{\beta_c}{2N^2}\frac{1}{L_t} .
\end{align}
As remarked earlier, at finite $\beta$ the lattice coupling is scheme dependent,
and we will choose to use the mean field improved coupling, replacing $\beta$ by 
$\beta_I=\beta \Braket{\frac{1}{N}\text{tr}(U_p)}$ in the above.
Once we have $T_c/(g^2N)$ for each of our $L_t$ values, we can take the continuum limit 
\begin{align}
\frac{T_c}{g^2N}(a)=\frac{T_c}{g^2N}(a=0)+c ag^2N+\cdots
\label{phase:couplingcontinuumextrapolation}
\end{align}
where the leading order correction is $\mathcal{O}(a)$ rather than  $\mathcal{O}(a^2)$ 
since, unlike the string tension or glueball mass, the lattice coupling is not a 
physical quantity.

We note that the errors on the values of $\beta_c$ and  $\beta_{I,c} $ are typically 
much smaller than on the $a\surd\sigma$ lattice values. However this greater 
accuracy is offset by the fact that these values are `further away' from the
continuum limit in that the leading correction is $O(a)$ rather than $O(a^2)$.
Moreover one would naively expect $T_c$ and $\sigma$ to be more closely correlated 
than $T_c$ and some lattice $g^2$, and so their ratio to be closer to its
continuum value. For this reason we will place more stress on our continuum
extrapolation of $T_c/\surd\sigma$ than on $T_c/g^2N$.

Finally we remark that we could equally well  express the critical temperature 
in units of the lightest scalar glueball 
mass $m_{0^+}$, by calculating this mass at each $\beta_c$ in the $T\simeq 0$ theory,
and extrapolating the resulting dimensionless ratio to the continuum limit.
However we do not do this here. Rather we simply obtain the continuum ratio 
$T_c/m_{0^+}$ from the continuum limit of  $m_{0^+}/\surd\sigma$ calculated in
\cite{Lau:aa}
and our extrapolated value of  $T_c/\surd\sigma$,
\begin{align}
\frac{T_c}{m_{0^+}}=\frac{T_c/\surd\sigma}{m_{0^+}/\surd\sigma} .
\end{align}

\subsection{Bulk transition}

Lattice gauge theories generally have some kind of `bulk' transition between 
the regions of strong and weak coupling, where the coupling expansion changes 
from powers of $\beta\propto 1/(ag^2)$ to powers of $1/\beta\propto ag^2$.
Since an extrapolation to the continuum limit, $\beta\to\infty$, is only 
plausible, {\it a priori}, if made using values obtained in the weak coupling
region, it is important to know where this bulk transition occurs. 

With the $SO(N)$ plaquette action, we find that the bulk transition seems 
to be characterised by the appearance of a very light excitation in the scalar 
glueball sector, with the rest of the glueball spectrum being essentially
unaffected.
Moreover we find that the visibility of this light excitation is sensitive
to the lattice volume and that as $N$ increases, we can use smaller volumes
to identify the bulk transition in this way. 
This is an interesting and unusual transition, which we will describe in
greater detail in our companion paper on the glueball spectrum
\cite{Lau:aa}.
For our present purposes, we only need to note that it provides an unambiguous 
way to identify the location of the bulk transition. 
We show the $\beta$ values corresponding to this bulk transition in 
Table~\ref{phase:tab:bulktransition}
together with the range of $L_t$ values for which the corresponding $\beta_c$
lie in the weak coupling region. We note that the transition moves to weaker 
coupling as $N$ decreases, making the weak coupling calculations more expensive
at small $N$.
This is why we have not performed $SO(3)$ calculations, which one can estimate 
would necessitate using $L_t > 10$ (and up to $L_t \sim 20$ to have a
useful lever arm for a continuum extrapolation).

To calculate the continuum limit of the deconfinement temperature, we shall use 
data corresponding to values in the weak coupling region, ignoring  the data from 
$L_t$ values that have $\beta_c$ values in the strong coupling region.
Occasionally, where the would-be susceptibility peak around $\beta_c$ overlaps with
this bulk transition, it may be grossly distorted by the very light
scalar excitation (which can also affect the winding flux tube spectrum)
and we are then unable to obtain a usefully precise value of  $\beta_c$.  

\subsection{$SO(4)$}

For $SO(4)$, the $\beta_c$ values for $L_t<5$ are in the strong coupling region 
whereas the $\beta_c$ values for $L_t>5$ are in the weak coupling region. The
deconfining transition for $L_t=5$ mixes with the bulk transition and we do not
attempt to extract corresponding values of $\beta_c$.  We give the corresponding 
values of $T_c/\surd\sigma$ in Table~\ref{phase:tab:so4continuum}. 

\begin{figure}
	\centering
  	\includegraphics[width=\textwidth]{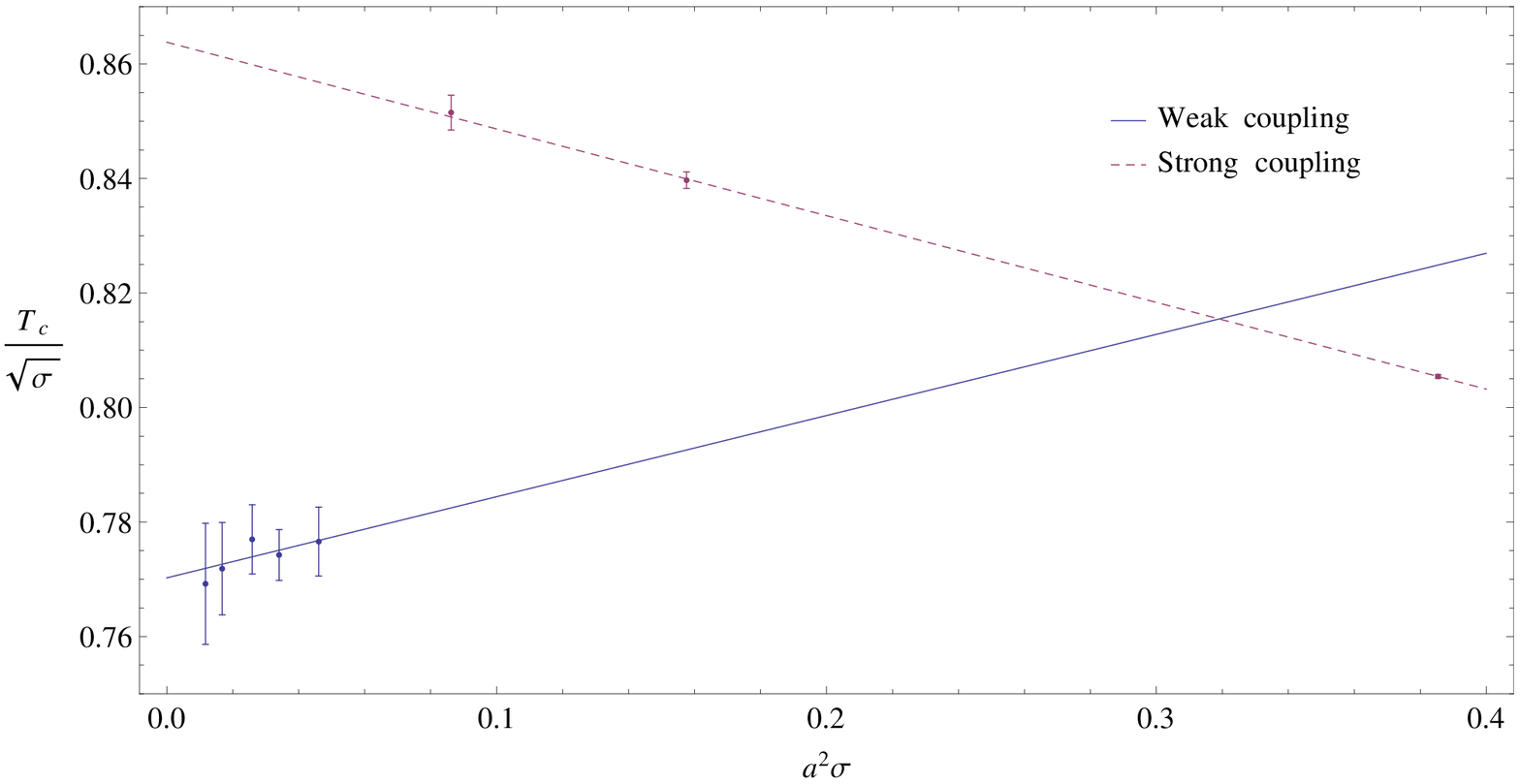} 
	\caption{Continuum extrapolation of $SO(4)$ deconfining temperature
          in units of the string tension.}
	\label{phase:fig:so4continuum}
\end{figure}

We extrapolate the values of $T_c/\surd\sigma$ to the continuum limit using 
eqn\eqref{phase:stringcontinuumextrapolation}.
We display the data and the fits in Figure~\ref{phase:fig:so4continuum}. 
There are two separate fits on display. The first is to the weak-coupling data, 
obtained on lattices with $L_t \geq 6$. This data shows very little
dependence on $a$ and the fit with just the leading $O(a^2)$ correction works 
well. This is no surprise because $a^2\sigma \ll 1$ for all the weak coupling data.
We obtain a continuum limit
\begin{align}
\frac{T_c}{\surd\sigma}(a=0) = 0.7702(88)  
\quad  \bar{\chi}^2_\text{dof}=0.12 \quad SO(4)\text{ (weak coupling)} .
\label{eqn:so4contTk}
\end{align}
The second fit is motivated by the fact that the three strong coupling values
obtained on lattices with $L_t\leq 4$, appear to lie on a stright line. A linear
fit as in  eqn\eqref{phase:stringcontinuumextrapolation} works well and provides 
us with what we dub a `strong coupling' continuum limit
\begin{align}
\frac{T_c}{\surd\sigma}(a=0) = 0.8638(21) 
\quad \bar{\chi}^2_\text{dof}=0.09 \quad SO(4)\text{ (strong coupling)} .
\end{align}
This linearity of the strong coupling data is unexpected and indeed bizarre. 
It may just be an accident, in which case our exercise is meaningless. 
However it may be that $a^2$ is small enough for the operator expansion
of the lattice action in powers of $a^2$ is viable even if the coupling expansion 
in powers of $1/\beta$ is not. If so one might speculate that this provides some 
kind of strong coupling continuum limit. In any case, the true continuum limit of
the $SO(4)$ theory is the one extracted from the weak coupling values in
eqn(\ref{eqn:so4contTk}).

Similarly, we can calculate the critical temperatures in units of the 't Hooft coupling.
The values of $T_c/(g^2N)$ are listed in Table~\ref{phase:tab:so4continuum}. 
We can plot $T_c/(g^2N)$ against $ag^2N$ and extrapolate to the continuum limit using 
eqn\eqref{phase:couplingcontinuumextrapolation}.
The continuum limit is
\begin{align}
&&\frac{T_c}{g^2N}(a=0)&=0.04567(43) 	&&\bar{\chi}^2_\text{dof}=2.17	
&&SO(4)\text{ (weak coupling)}
\end{align}
where we need to drop the $L_t=6$ point from the fit in order to obtain a reasonable 
$\bar{\chi}^2_\text{dof}$ with just a leading order weak coupling correction. (Note 
that the strong coupling values do not fit onto a linear extrapolation in $1/\beta_I$, 
which is of course as expected.) 

Finally, we express the critical temperature in units of the lightest scalar glueball 
mass $M_{0^+}$. We use the continuum value of $M_{0^+}/\surd\sigma$ calculated in
\cite{Lau:aa} 
with our above continuum value of  $T_c/\surd\sigma$ to obtain 
\begin{align}
\frac{T_c}{M_{0^+}}(a=0)=0.2293(30) \qquad SO(4)\text{ (weak coupling)} .
\end{align}

\subsection{$SO(5)$ and $SO(6)$}

\begin{figure}
	\centering
  	\includegraphics[width=\textwidth]{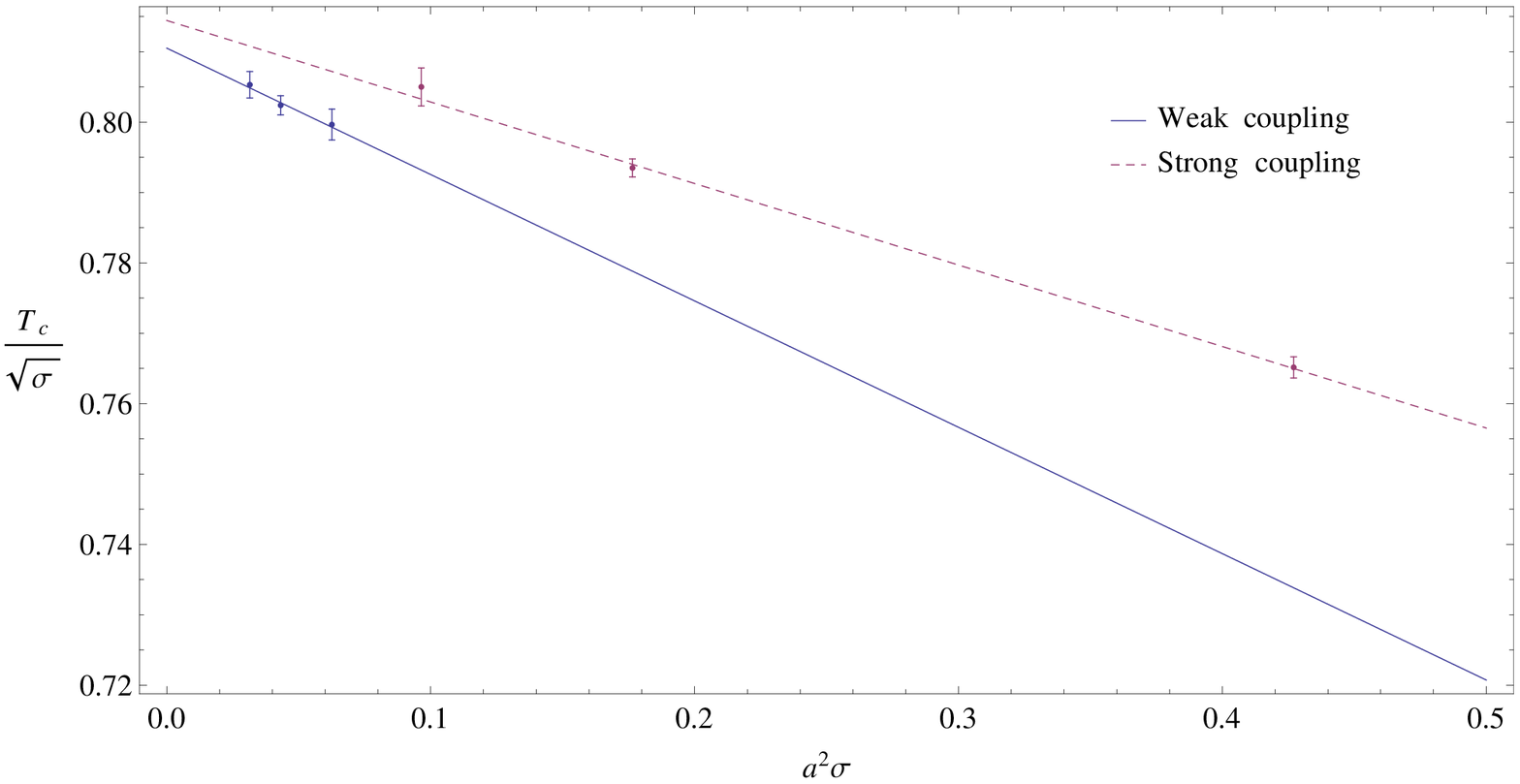} 
	\caption{Continuum extrapolation of $SO(6)$ deconfining temperature
          in units of the string tension.}
	\label{phase:fig:so6continuum}
\end{figure}

For both $SO(5)$ and $SO(6)$, the $\beta_c$ values for $L_t\geq5$ are in the weak 
coupling region and so can be used for a continuum extrapolation. 
To obtain the critical temperature in string tension units $T_c/\surd\sigma$, we 
calculate the string tension at each $\beta_c$ as in $SO(4)$.
We list the resulting values for $SO(5)$ in Table~\ref{phase:tab:so5continuum} and for 
$SO(6)$ in Table~\ref{phase:tab:so6continuum}. 
We display the continuum extrapolation for $SO(6)$ in Figure~\ref{phase:fig:so6continuum}. 

In the case of $SO(5)$, unlike $SO(4)$, there were difficulties in using a linear extrapolation 
in the weak coupling region due to peculiar variation in the value of $T_c/\surd\sigma$.
To obtain a good fit we had to drop the two smallest $L_t$ points in the weak coupling
region. (Context: for no other $N$ did we need to drop any weak coupling points.) 
$SO(5)$ is the largest $SO(N)$ group for which the transition is second order and it might
be that this is behind this atypical behaviour.
The continuum limit from within the weak coupling region is
\begin{align}
&&\frac{T_c}{\surd\sigma}(a=0)&=0.7963(114), &&\bar{\chi}^2_\text{dof}=0.003	
&&SO(5) .
\end{align}
We also note that the strong coupling values fit less well with a linear extrapolation than 
they did for $SO(4)$, giving a strong coupling `continuum' extrapolation of 
$T_c/\surd\sigma=0.783(4)$ with a mediocre $\bar{\chi}^2_\text{dof}=2.78$.

$SO(6)$ is the smallest group for which the transition is first order. The continuum limit 
taken from data within the weak coupling region is
\begin{align}
&&\frac{T_c}{\surd\sigma}(a=0)&=0.8105(42), 	&&\bar{\chi}^2_\text{dof}=0.16	&&SO(6) .
\end{align}
There is also a good linear extrapolation using the strong coupling values that
gives $T_c/\surd\sigma=0.8144(20)$ with $\bar{\chi}^2_\text{dof}=0.59$. We note
that here the strong coupling extrapolation is consistent with the true weak-coupling 
continuum limit.

We can also calculate the critical temperatures in units of the coupling. $T_c/(g^2N)$.
The values are listed in Table~\ref{phase:tab:so5continuum} and
Table~\ref{phase:tab:so6continuum}. 
We can then plot $T_c/(g^2N)$ against $ag^2N$ and extrapolate to the continuum limit using 
eqn\eqref{phase:couplingcontinuumextrapolation}.
The continuum limits, from within the weak coupling regions, are
\begin{equation}
\left.\frac{T_c}{g^2N} \right|_{a = 0}
=
\begin{cases}
  0.05544(92) 	\quad \bar{\chi}^2_\text{dof}=0.05	\quad SO(5) &\\ 
  0.05996(19)	\quad \bar{\chi}^2_\text{dof}=0.53	\quad SO(6) & 
\end{cases}
.
\end{equation}
In the case of the $SO(5)$ data, the points seem to lie on a smooth curve and do not
exhibit the peculiar variation seen in the corresponding $T_c/\surd\sigma$ values.

Finally, we can calculate the critical temperatures in units of the lightest scalar 
glueball. Using the values of $M_{0^+}/\surd\sigma$ from 
\cite{Lau:aa}
we obtain
\begin{equation}
\left.\frac{T_c}{M_{0^+}} \right|_{a = 0}
=
\begin{cases}
  0.2244(33) \quad SO(5)&\\
  0.2232(14) \quad SO(6)&
\end{cases}
.
\end{equation}

\subsection{$SO(7)$, $SO(8)$, $SO(9)$, $SO(12)$, and $SO(16)$}

For $SO(7)$ and $SO(8)$ the $\beta_c$ values are in the weak coupling region for $L_t\geq4$, 
and for $SO(9)$, $SO(12)$, and $SO(16)$ they are in the weak coupling region for $L_t\geq3$.

We list the critical temperature values in string tension units $T_c/\surd\sigma$ for 
these groups in Tables~\ref{phase:tab:so7continuum}, \ref{phase:tab:so8continuum}, 
\ref{phase:tab:so9continuum},  \ref{phase:tab:so12continuum}, and
\ref{phase:tab:so16continuum}.
The continuum limits are
\begin{equation}
\left.\frac{T_c}{\surd\sigma} \right|_{a = 0}
=
\begin{cases}
  0.8351(38) 	 \quad \bar{\chi}^2_\text{dof}=0.98	 \quad SO(7)& \\
  0.8418(39) 	 \quad \bar{\chi}^2_\text{dof}=0.05	 \quad SO(8)& \\
  0.8515(14) 	 \quad \bar{\chi}^2_\text{dof}=0.30       \quad SO(9)& \\
  0.8642(38) 	 \quad \bar{\chi}^2_\text{dof}=0.02	 \quad SO(12)& \\
  0.8780(38) 	 \quad \bar{\chi}^2_\text{dof}=0.15	 \quad SO(16)& 
\end{cases}
.
\end{equation}
We note that all these fits are very good.

Similarly, we can calculate the critical temperature in units of the coupling, as listed 
in Tables~\ref{phase:tab:so7continuum}, \ref{phase:tab:so8continuum}, 
\ref{phase:tab:so9continuum},  \ref{phase:tab:so12continuum}, and  \ref{phase:tab:so16continuum}. 
We can then plot $T_c/(g^2N)$ against $ag^2N$ and extrapolate to the continuum limit 
using eqn\eqref{phase:couplingcontinuumextrapolation}. These plots have very similar forms 
to the corresponding $T_c/\surd\sigma$ plots.
The continuum limits are
\begin{equation}
\left.\frac{T_c}{g^2N} \right|_{a = 0}
=
\begin{cases}
  0.06478(18) 	 \quad  &\bar{\chi}^2_{dof}=4.01	 \quad SO(7)  \\
  0.06809(16)	 \quad  &\bar{\chi}^2_{dof}=0.00	 \quad SO(8)  \\
  0.07043(7) 	 \quad  &\bar{\chi}^2_{dof}=0.10	 \quad SO(9) \\
  0.07552(14) 	 \quad  &\bar{\chi}^2_{dof}=0.63	 \quad SO(12)  \\
  0.07947(17) 	 \quad  &\bar{\chi}^2_{dof}=0.85	 \quad SO(16) 
\end{cases}
.
\end{equation}
We can see that these continuum extrapolations are mostly good. (Given there is only one
degree of freedom, the $SO(7)$ fit is not unacceptable.) 

Finally, we can calculate the critical temperatures in units of the lightest scalar 
glueball mass $M_{0^+}$. Using the values for $N=7,8,12,16$ calculated in
\cite{Lau:aa}
(there is no calculation for $N=9$) we find
\begin{equation}
\left.\frac{T_c}{M_{0^+}} \right|_{a = 0}
=
\begin{cases}
  0.2234(12)   \quad SO(7) & \\
  0.2224(15)   \quad SO(8) & \\
  0.2217(18)   \quad SO(12) & \\
  0.2220(22)   \quad SO(16) & 
\end{cases}
.
\end{equation}

\section{Results: large-$N$ limits}
\label{sec:largen}

\subsection{Deconfining temperature}

In contrast to $SU(N)$, the leading large-$N$ correction for $SO(N)$ gauge theories
is expected to be  $\mathcal{O}(1/N)$. So we expect
\begin{align}
\left.\frac{T_c}{\mu} \right|_{SO(N)}
\stackrel{N\to\infty}{=}\left.\frac{T_c}{\mu} \right|_{SO(\infty)}
+ \frac{c}{N} + \cdots
\end{align}
with $\mu$ a physical mass scale such as $\surd\sigma$, $m_{0^+}$ or $g^2N$.

Since one of our aims is to compare the values of $T_c$ for $SO(2N)$ and $SO(2N+1)$
gauge theories, it would be useful to have an estimate of $T_c$ in  $SO(3)$.
Since the $SO(3)$ and $SU(2)$ groups have the same Lie algebra, it is plausible
to assume that they share the same value of $T_c/M_{0^+}$, where  $M_{0^+}$ is the
mass of the lightest scalar glueball.
We have to be more careful with  $T_c/\surd\sigma$ because the fundamental string
tension in $SO(3)$ corresponds to the adjoint in $SU(2)$.
Now, we know that in $SU(2)$ $T_c/\surd\sigma=1.1238(88)$
\cite{Jack-Liddle:2008xr}
We also know that in  $SU(2)$  $M_{0^+}/\surd\sigma=4.7367(55)$ 
\cite{Athenodorou:aa}
and in $SO(3)$ $M_{0^+}/\surd\sigma=2.980(24)$
\cite{Lau:aa}.
From the ratio of these two numbers we extract an estimate of the ratio of
fundamental string tensions in $SU(2)$ and $SO(3)$.
All this implies that the $SO(3)$ continuum deconfining temperature 
in units of the string tension is
\begin{align}
&&\frac{T_c}{\surd\sigma}&=0.7072(80) 	&&SO(3).&&
\end{align}
We also know the $SO(3)$ string tension $\surd\sigma/(g^2N)=0.04576(36)$
\cite{Lau:aa},
which tells us that
\begin{align}
&&\frac{T_c}{g^2N}&=0.03236(45) 	&&SO(3).&&
\end{align}
Finally, we can also infer from the above that the $SO(3)$ continuum
deconfining temperature in units of the lightest scalar glueball mass is
\begin{align}
&&\frac{T_c}{M_{0^+}}&=0.2373(33) 	&&SO(3).&&
\end{align}

\begin{figure}
	\centering
  	\includegraphics[width=\textwidth]{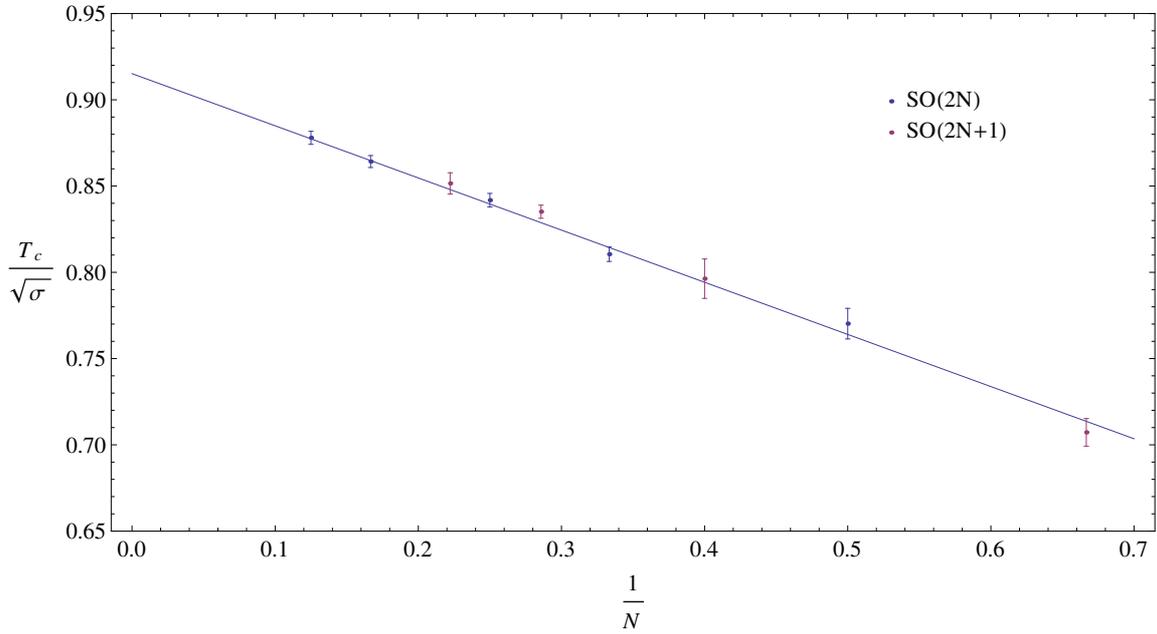} 
	\caption{Large-$N$ extrapolation of $SO(2N)$ deconfining temperature
          in units of the string tension. Plotted points include both $SO(2N)$
          and $SO(2N+1)$.}
	\label{phase:fig:sonwithso2nfitstring}
\end{figure}

We list the $SO(N)$ deconfining temperatures in string tension units in
Table~\ref{phase:tab:soncontinuum}.
We begin by applying a linear fit in $1/N$ to just the $SO(2N)$ values. 
We do so for two reasons.
Firstly, we intend to compare this limit to the $SU(N)$ large-$N$ limit 
motivated by the large-$N$ orbifold equivalence.
Secondly, $SO(2N+1)$ has a different centre to $SO(2N)$, so  the deconfinement 
properties might differ between the two sets of gauge theories and it is
interesting to see if this is the case. 
In Figure~\ref{phase:fig:sonwithso2nfitstring} we plot all our values of 
${T_c}/{\surd\sigma}$, including that inferred for $SO(3)$, against $1/N$, 
and we also show the best leading-order fit to just the  $SO(2N)$ values.
We see that the linear fit is very good. We also see that values for the
$SO(2N+1)$ groups are consistent with lying on this fit. Indeed if we
take all the values, including $SO(3)$, we obtain a very similar best fit:
\begin{equation}
\left.\frac{T_c}{\surd\sigma} \right|_{N \rightarrow \infty}
=
\begin{cases} 
0.9152(48)  \quad \bar{\chi}^2_\text{dof}=0.58 \quad SO(2N\geq 4) & \\
0.9231(45)  \quad \bar{\chi}^2_\text{dof}=0.52 \quad SO(2N+1\geq 3) & \\
0.9194(33)  \quad \bar{\chi}^2_\text{dof}=0.82 \quad SO(N\geq 3) &
\end{cases}
.
\end{equation}
We conclude that at our level of accuracy there is no evidence 
for any difference in the way $T_c/\surd\sigma$ varies with $N$ in 
$SO(2N)$ and  $SO(2N+1)$ gauge theories: the lack of a center symmetry
in the latter appears to play no role.   
We also observe that the groups with a second order transition,
$SO(N\leq 5)$, fall nicely on the smooth curve that describes the
$N$-dependence of the first-order transitions, $SO(N\geq 6)$.

We can repeat the above, replacing $\surd\sigma$ by the 't Hooft coupling $g^2N$.
We list the $SO(N)$ deconfining temperatures in units of $g^2N$ in 
Table~\ref{phase:tab:soncontinuum}.
To  fit to all the values of $N$, we need to include an additional $O(1/N^2)$
correction and, to avoid a systematic bias, we do the same for our separate fits 
to odd and even $N$. We find:
\begin{equation}
\left.\frac{T_c}{g^2N} \right|_{N \rightarrow \infty}
=
\begin{cases} 
0.09139(49) \quad \bar{\chi}^2_\text{dof}=2.53 \quad SO(2N\geq 4)&\\
0.09039(63) \quad \bar{\chi}^2_\text{dof}=0.43 \quad SO(2N+1\geq 3)&\\
0.09160(35) \quad \bar{\chi}^2_\text{dof}=1.45 \quad SO(N\geq 3 )&
\end{cases}
.
\label{eqn:ggNlargeN}
\end{equation}
We see that values of $T_c/(g^2N)$ obtained for the $SO(2N+1)$ groups are 
consistent with those for $SO(2N)$.

\begin{figure}
	\centering
  	\includegraphics[width=\textwidth]{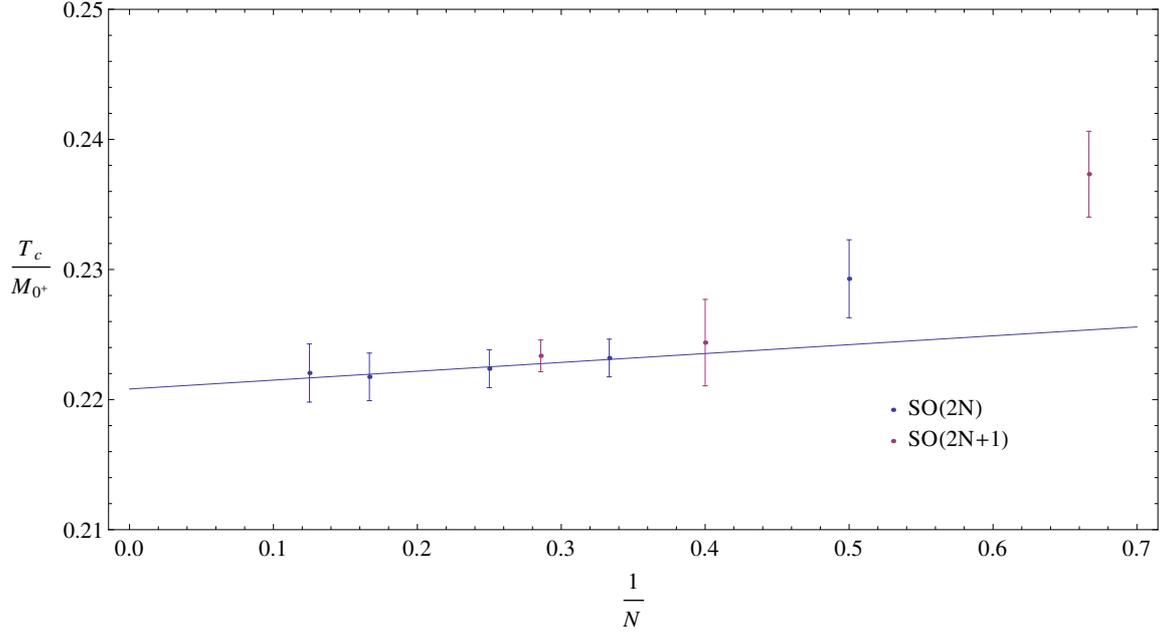} 
	\caption{Large-$N$ extrapolation of $SO(2N)$ deconfining temperature in units
          of the lightest scalar glueball mass.  Plotted points include both $SO(2N)$
          and $SO(2N+1)$.}
	\label{phase:fig:sonwithso2nfitglueball}
\end{figure}

Finally, we list the $SO(N)$ deconfining temperatures in units of the lightest 
scalar glueball mass in Table~\ref{phase:tab:soncontinuum} and plot
these values in Figure~\ref{phase:fig:sonwithso2nfitglueball}.
We show a leading-order fit to $SO(2N)$ for $2N\geq 6$ since the data 
(if one pays attention to the $SO(3)$ value) indicates the need for a higher 
order correction at the lower values of $N$.
We do not fit odd $N$ separately because we do not have available a glueball 
mass for $SO(9)$, and so the number of odd $N$ values is too small to fit. 
We also perform fits with an additional  $O(1/N^2)$ correction to 
both $SO(2N\geq 4)$ and to all our values, $SO(N\geq 3)$.
Altogether, these fits give:
\begin{equation}
\left.\frac{T_c}{M_{0^+}} \right|_{N \rightarrow \infty}
=
\begin{cases} 
0.2209(28) \quad \bar{\chi}^2_\text{dof}=0.03 \quad SO(2N\geq 6)&\\
0.2189(23) \quad \bar{\chi}^2_\text{dof}=0.51 \quad SO(2N\geq 4)&\\
0.2230(39) \quad \bar{\chi}^2_\text{dof}=0.10 \quad SO(N\geq 3)&
\end{cases}
.
\end{equation}
We note that the linear fit is particularly flat compared to the linear fit
in Figure~\ref{phase:fig:sonwithso2nfitstring} and the one to ${T_c}/{g^2N}$.
Finally, we again see evidence that the $SO(2N)$ and  $SO(2N+1)$ values form a single
smooth series.

\subsection{Large-$N$ scaling}
\label{subsec:Nscaling}

We have assumed throughout that the large-$N$ limit requires keeping $g^2N$ fixed
and that the leading correction is $O(1/N)$, guided by the all-orders analysis
of diagrams
\cite{Lovelace:1982fk}.
Certainly, keeping $g^2N$ fixed is necessary if one wants to obtain an $SO(\infty)$ 
theory that is perturbative (and asymptotically free) at short distances. Here we 
ask whether our non-perturbative calculations support these assumptions.

Without assuming $g^2N$ scaling, we can test for the power of the leading 
correction by fitting 
\begin{align}
  &&\left.\frac{T_c}{\surd\sigma} \right|_{SO(N \rightarrow \infty)}
  = c_0 + \frac{c_1}{N^\alpha}&& SO(N\geq 3) .
\end{align}
We find that $\alpha = 1.13 \pm 0.14$. If we assume $g^2N$ scaling then a similar 
analysis for $T_c/g^2N$, over the range $N\geq 7$ where we can  get a good fit, gives 
$\alpha = 0.88 \pm 0.16$. So if the power of the correction is an integer, then
this confirms that it must be $O(1/N)$.

It is amusing to see if our results also demand that $g^2N$ should be kept constant.
We can fit 
\begin{align}
  &&\left.\frac{T_c}{g^2N^\gamma} \right|_{SO(N \rightarrow \infty)}
  = c_0 + \frac{c_1}{N}&& SO(N\geq 7)&& 
\end{align}
and doing so we find a tight constraint $\gamma=1.020\pm 0.024$, just as expected.

\section{Comparison of $SO(N)$ and $SU(N)$ deconfining temperatures}
\label{sec:comparison}

\subsection{$SO(4)\sim SU(2)\times SU(2)$}

We know that $SO(4)$ and $SU(2)\times SU(2)$ share a common Lie algebra so it is
interesting to see if they have the same deconfining temperatures and transitions.
We have seen that the $SO(4)$ deconfining phase transition is second order, 
just like $SU(2)$, and (within large errors) they appear to share the same 
critical exponents.
Now, we expect that the fundamental flux of $SO(4)$ contains the fundamental flux
of both $SU(2)$ groups from the product group $SU(2)\times SU(2)$, so that
\begin{align}
\left.\sigma \right|_{su2\times su2}=2\left.\sigma \right|_{su2} .
\end{align}
Hence, we expect that
\begin{align}
\left.\frac{T_c}{\surd\sigma} \right|_{so4}
=\left.\frac{T_c}{\surd\sigma} \right|_{su2\times su2}
=\frac{1}{\sqrt2}\left.\frac{T_c}{\surd\sigma} \right|_{su2} .
\label{phase:so4su2}
\end{align}
We know that the $SU(2)$ deconfining temperature is $T_c/\surd\sigma=1.1238(88)$ 
\cite{Jack-Liddle:2008xr} 
so that we can compare this to our value for $SO(4)$:
\begin{align}
&&&&\frac{T_c}{\surd\sigma} 					&=0.7702(88)&SO(4)&&&&\nonumber\\
&&&&\frac{1}{\sqrt2}\frac{T_c}{\surd\sigma} 	&=0.7946(62)&SU(2)&.&&& 
\end{align}
We see that these values are within about $2.25\sigma$ of each other which we consider 
to be reasonable agreement.

\subsection{$SO(6)\sim SU(4)$}

$SO(6)$ and $SU(4)$ also share a common Lie algebra so it is also interesting to 
compare their deconfining transitions.
We have seen that $SO(6)$ is first order, but weakly so, and this is also
the case for $SU(4)$
\cite{Jack-Liddle:2008xr,Pepe:2007}. 
As we discussed earlier, the $SO(6)$ fundamental string tension is equivalent 
to the $SU(4)$ $k=2$ anti-symmetric string tension so what we may expect is
\begin{align}
\left.\frac{T_c}{\surd\sigma_f} \right|_{so6}=\left.\frac{T_c}{\surd\sigma_{2A}} \right|_{su4} .
\label{phase:so6su4}
\end{align}
Hence, to compare between the $SO(6)$ and $SU(4)$ deconfining temperatures measured 
in units of the fundamental string tension, we need the ratio of the $k=2A$ 
and fundamental string tensions in $SU(4)$, and this has been calculated to be 
$\left.\sigma_{2A}/\sigma_{f}\right|_{su4}=1.355(9)$ in
\cite{Bringoltz2008429}. 
We also know from 
\cite{Jack-Liddle:2008xr},
that the $SU(4)$ deconfining temperature is $T_c/\surd\sigma_f=0.9572(39)$ so that 
\begin{align}
&&&&\frac{T_c}{\surd\sigma_{f}} =0.8105(42)&&SO(6)&&&&\nonumber\\
&&&&\frac{T_c}{\surd\sigma_{2A}}=0.8223(61)&&SU(4)&.&&& 
\end{align}
We see that these values are in agreement, being within $1.5\sigma$ of each other.

\subsection{Large-$N$ (orbifold) equivalence}

As remarked earlier, the existence of an orbifold projection from $SO(2N)$ to 
$SU(N)$ gauge theories means that they should have the same large-$N$ limit
and in particular the same deconfining temperature in that limit when expressed 
in physical units. In addition we have the diagrammatic planar equivalence 
of $SO(N)$ and $SU(N)$.

\begin{figure}
	\centering
  	\includegraphics[width=\textwidth]{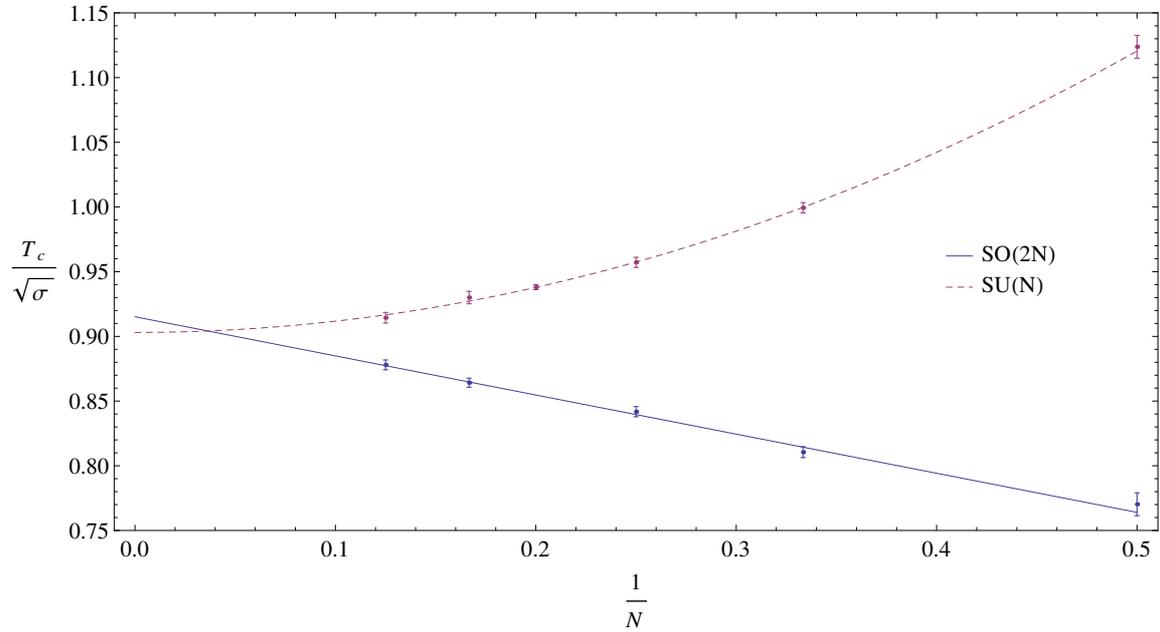} 
	\caption{Large-$N$ extrapolations of $SO(2N)$ and $SU(N)$ deconfining
          temperatures in units of the string tension.}
	\label{phase:fig:largenstring}
\end{figure}

We list the $SO(2N)$ and $SU(N)$ 
\cite{Jack-Liddle:2008xr} 
continuum values of $T_c/\surd\sigma$ in Table~\ref{phase:tab:sonsun}.
We display the corresponding large-$N$ extrapolations in 
Figure~\ref{phase:fig:largenstring}. The two large-$N$ limits are
\begin{equation}
\frac{T_c}{\surd\sigma}
=
\begin{cases}
  0.9152(48) \quad SO(2N \rightarrow \infty)& \\
  0.9030(29) \quad SU(N \rightarrow \infty)& 
\end{cases}
 .
\end{equation}
We see that these two values are within $2\sigma$ of each other, which is 
consistent with the hypothesis that they are equal. As for the planar
equivalence, we recall that fitting $SO(N\geq 3)$ with a fit that is linear 
in $1/N$ gives $T_c/\surd\sigma =0.9194(33)$ at $N=\infty$, and this is a
less comfortable $\sim 4\sigma$ from the $SU(\infty)$ value.

Similarly, we list the $SO(2N)$ and $SU(N)$ 
\cite{Liddle:2006aa} 
continuum values of $T_c/(g^2N)$ in Table~\ref{phase:tab:sonsun}.
To compare the $SO(2N)$ and $SU(N)$ values, we need to rescale the $SO(N)$ values 
to $SO(2N)$ values by doubling them (as we did in some earlier figures).
This is due to the large-$N$ coupling matching $g^2_{SU(N)}N=g^2_{SO(2N)}N$,
so that the large-$N$ limit is
\begin{align}
\lim_{N \to \infty}\frac{T_c}{g^2_{SU(N)}N}=
\lim_{N \to \infty}\frac{T_c}{g^2_{SO(2N)}N}
=\lim_{N \to \infty}2\frac{T_c}{g^2_{SO(2N)}2N} .
\end{align} 
The two large-$N$ limits are
\begin{equation}
\frac{T_c}{g^2N}
=
\begin{cases}
  0.1828(10) \quad SO(2N \rightarrow \infty) & \\
  0.1852(8) \quad SU(N \rightarrow \infty) &
\end{cases}
\end{equation}
while a fit to $SO(N\geq 3)$ gives ${T_c}/{g^2N}=0.1832(7)$.
We see that these two values are no more than $\sim 2\sigma$ from each other.

\begin{figure}
	\centering
  	\includegraphics[width=\textwidth]{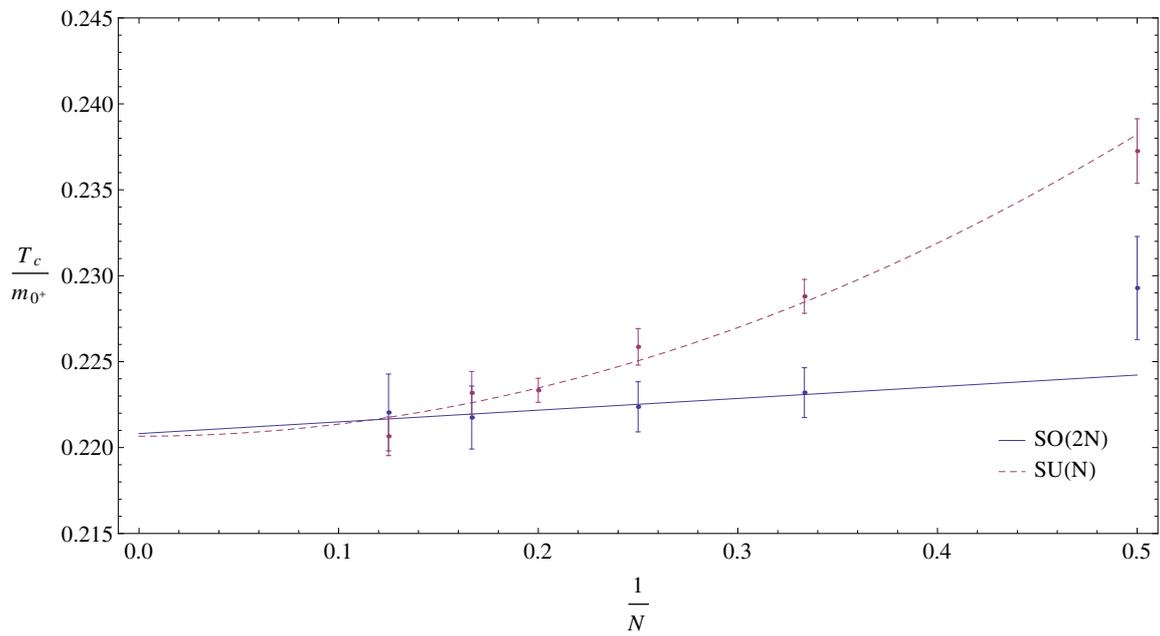} 
	\caption{Large-$N$ extrapolation of $SO(2N)$ and $SU(N)$ deconfining
          temperatures in units of the lightest scalar glueball mass.}
	\label{phase:fig:largenglueball}
\end{figure}

Finally, we list the $SO(2N)$ and $SU(N)$ 
\cite{Jack-Liddle:2008xr,Lucini:2002oz,Athenodorou:aa} 
values of  $T_c/M_{0^+}$  Table~\ref{phase:tab:sonsun}.
We display the two large-$N$ extrapolations in Figure~\ref{phase:fig:largenglueball}. 
The two large-$N$ limits obtained from these leading order fits are
\begin{equation}
\frac{T_c}{M_{0^+}}
=
\begin{cases}
  0.2209(28) \quad &SO(2N \rightarrow \infty)\\
  0.2207(6) \quad  &SU(N \rightarrow \infty)
\end{cases}
.
\end{equation}
and a higher order fit to all $SO(N\geq 3)$ gives a value $0.2230(39)$.
We see that all these  values agree very well.

\section{Conclusions}
\label{sec:conclusions}

In this paper we identified a finite temperature transition in $D=2+1$ 
$SO(N)$ gauge theories for $N=4,5,6,7,8,9,12,16$. For $N=4,5$ the transition
appears to be second order, while for $N\geq 6$ it appears to be first order.
We did not attempt to calculate $T_c$ for $SO(3)$ because the inconvenient location 
of the `bulk' transition would have made it computationally expensive.
However, the close connection between $SO(3)$ and $SU(2)$ makes one confident
that there is a deconfining transition in $SO(3)$, and we have used the $SU(2)$ 
value of $T_c$ to provide an estimate for the $SO(3)$ value.

This transition appears to have all the characteristics of a deconfining
transition for both even and odd $N$, and appears to be a phase transition
rather than a cross-over. We gave some arguments, and provided some evidence,
that $SO(N)$ gauge theories are indeed confining at low $T$, and that this 
is the case not just for even $N$ but also for odd $N$ where the centre 
is trivial.
  
Our calculations were performed on a lattice in a finite volume, but our 
final results for the deconfining temperature, $T_c$, are for the continuum 
theory in an infinite volume. (Achieved through extrapolation, of course.)
We find that dimensionless ratios such as $T_c/\surd\sigma$, where $\sigma$
is the zero-temperature confining string tension, fall on a single sequence that
can be interpolated by a smooth function of $N$ for $N\geq 3$ (with the value for
$SO(3)$ being inferred from the value in $SU(2)$). That is to say, we can
think of the $N$-dependence of non-Abelian $SO(N)$ gauge theories as a continuous 
function of $N$ for all $N$. In particular there is no evidence that even and odd 
$N$ fall on two separate (even if converging) branches.

Somewhat remarkably we find that a simple leading-order large-$N$ expression 
suffices to fit all our calculated  values of  $T_c/\surd\sigma$:
\begin{align}
\frac{T_c}{\surd\sigma} = 0.9194(33) - \frac{0.620(28)}{N}  \quad ; \quad N\geq 3 .
\end{align}
Such a `precocious' large-$N$ scaling seems the norm for physical mass ratios 
in both $SO(N)$
\cite{Lau:aa,Bursa:2013lq}
and $SU(N)$ 
\cite{Teper:1998rw,Lucini:2002oz,Athenodorou:aa}
gauge theories.
Even more striking is how weakly the ratio ${T_c}/{M_{0^+}}$ depends on $N$, 
as we see in Figure~\ref{phase:fig:largenglueball} and as
highlighted by the smallness of the leading correction in our higher order fit
\begin{align}
\frac{T_c}{M_{0^+}} = 0.2230(39) - \frac{0.033(45)}{N} 
+ \frac{0.23(12)}{N^2}  \quad ; \quad N\geq 3 .
\end{align}
A possible explanation for this weak $N$-dependence is discussed in
\cite{Athenodorou:2015aa}.

As an aside, we remark that our results, as described in Section~\ref{subsec:Nscaling},
provide a non-perturbative confirmation of the expected large-$N$ scaling:
$g^2N$ fixed, and $O(1/N)$ leading corrections as $N\to\infty$. 

As another (much less expected) aside, we recall that our values of ${T_c}/{\surd\sigma}$ 
on the strong coupling side of the `bulk' transition also appeared to extrapolate
to $a=0$ with a simple $O(a^2)$ correction. At small $N$ this `strong coupling
continuum limit' was very different from the true weak coupling continuum
limit, but at larger $N$ they became consistent. This unexpected and bizarre
behaviour strongly suggests that our interpretation of the `bulk' transition 
as a simple strong-to-weak coupling transition is too naive.

Since $SU(N)$ gauge theories can be orbifold projected from $SO(2N)$, we expect
them to have the same physics at large $N$
\cite{Unsal:2006kx}, 
and indeed we find that the values of $T_c$ extrapolated to $N=\infty$ are 
consistent with being equal, at the $\sim 1\%$ level. We obtain a similar
confirmation of the large-$N$ planar equivalence between $SO(N)$ and $SU(N)$ 
gauge theories. In addition we find
that the values of $T_c$ in $SO(4)$ and $SO(6)$ are consistent with those
in $SU(2)\times SU(2)$ and $SU(4)$ respectively, indicating that for this
physics at least the differences in group global structure are not important: 
theories with the same Lie algebra appear to possess the same value of $T_c$.

\section*{Acknowledgements}

Our interest in this project was originally motivated by Aleksey
Cherman and Francis Bursa. We acknowledge, in particular, the key
contributions made by Francis Bursa to this project in its early stages.
The project originated in a number of discussions during the 2011 Workshop on 
`Large-N Gauge Theories'  at the Galileo Galilei Institute in Florence,
and we are grateful to the Institute for providing such an ideal environment 
within which to begin collaborations.
RL has been supported by an STFC STEP award under grant ST/J500641/1, 
and MT acknowledges partial support under STFC grant ST/L000474/1.
The numerical computations were carried out on EPSRC, STFC and Oxford funded 
computers in Oxford Theoretical Physics.

\clearpage

\appendix

\section{Tables}
\label{app:data}
\vspace{3cm}
\begin{table}[h]
\centering
\begin{tabular}{ |cccc|}
  \hline
  	Data Fit 	& $\beta_c$ & $\chi(\beta_c)$ & $\bar{\chi}^2_\text{dof}$\\
  \hline
Reweighting	&8.493(10)	&25.40(30) 	& n/a\\
Gaussian	&8.500(11)	&25.66(40)	& 0.62\\
Logistic	&8.500(6)	&25.71(41)	& 0.64\\
  \hline
\end{tabular}
	\caption{Comparisons between reweighting, Gaussian fits, and logistic fits to obtain $\beta_c$ and $\chi(\beta_c)$ on a $40^24$ volume in $SO(4)$.}
	\label{phase:tab:so4curvefitcomparison}
\end{table}

\vspace{2cm}

\begin{table}[h]
\centering
\begin{minipage}[t]{0.4\textwidth}
\vspace{0pt}
\begin{tabular}{|cll|}
  \hline
  	$L_s^2L_t$ 	& $\beta_c$ & $\chi_{\left| \overline{l_P}\right|}$\\
  \hline
$20^2 2$	&6.4748(4) 	&10.77(4)	\\
$24^2 2$	&6.4771(4)	&13.64(5)	\\
$28^2 2$	&6.4788(3) 	&16.80(9)	\\
$32^2 2$	&6.4797(3) 	&20.16(8)	\\
$36^2 2$	&6.4813(4) 	&23.65(12)	\\
$40^2 2$	&6.4819(3) 	&27.58(14)	\\
$48^2 2$	&6.4822(4) 	&35.15(43)	\\
$56^2 2$	&6.4840(4) 	&44.42(63)	\\
$60^2 2$	&6.4850(4) 	&49.92(89)	\\
$80^2 2$	&6.4853(4) 	&78.57(132)	\\
  \hline
\end{tabular}
\end{minipage}
\begin{minipage}[t]{0.4\textwidth}
\vspace{0pt}
\begin{tabular}{|cll|}
  \hline
  	$L_s^2L_t$ 	& $\beta_c$ & $\chi_{\left| \overline{l_P}\right|}$\\
  \hline
$32^2 3$	&7.534(3) 	&19.60(14)	\\
$36^2 3$	&7.538(3)	&23.22(16)	\\
$40^2 3$	&7.539(1) 	&26.37(25)	\\
$44^2 3$	&7.545(2) 	&31.09(38)	\\
$48^2 3$	&7.546(3) 	&35.56(38)	\\
$52^2 3$	&7.552(2) 	&40.58(53)	\\
$66^2 3$	&7.552(2) 	&58.75(131)	\\
$80^2 3$	&7.555(3) 	&81.02(193)	\\
$90^2 3$	&7.557(3) 	&95.80(171)	\\
  \hline
$40^2 4$	&8.493(10) 	&25.40(30)	\\
$48^2 4$	&8.501(6)	&33.88(56)	\\
$56^2 4$	&8.509(8) 	&42.02(93)	\\
$64^2 4$	&8.526(7) 	&51.67(120)	\\
$72^2 4$	&8.520(3) 	&59.46(140)	\\
$80^2 4$	&8.535(6) 	&66.44(173)	\\
$88^2 4$	&8.545(8) 	&75.23(296)	\\
  \hline
\end{tabular}
\end{minipage}
	\caption{$\beta_c$ and $\chi_{\left| \overline{l_P}\right|}$ in $SO(4)$ for $L_t\leq 4$ on various volumes.}
	\label{phase:tab:so4susceptibilitydatapart1}
\end{table}

\begin{table}
\centering
\begin{minipage}[t]{0.4\textwidth}
\vspace{0pt}
\begin{tabular}{|cll|}
  \hline
  	$L_s^2L_t$ 	& $\beta_c$ & $\chi_{\left| \overline{l_P}\right|}$\\
  \hline
$36^2 6$	&11.110(31) &20.68(18)	\\
$48^2 6$	&10.924(17)	&28.87(36)	\\
$60^2 6$	&10.861(9) 	&38.76(53)	\\
$72^2 6$	&10.837(14) &47.33(83)	\\
$84^2 6$	&10.809(21) &57.82(125)	\\
$96^2 6$	&10.824(14) &70.28(199)	\\
$120^2 6$	&10.835(8) 	&95.57(321)	\\
  \hline
$42^2 7$	&12.685(45) &29.41(38)	\\
$56^2 7$	&12.494(22)	&41.46(64)	\\
$70^2 7$	&12.397(12) &54.24(94)	\\
$84^2 7$	&12.303(12) &68.11(111)	\\
$98^2 7$	&12.224(12) &81.74(172)	\\
$112^2 7$	&12.261(19) &100.03(247)	\\
$126^2 7$	&12.274(19) &117.10(302)	\\
  \hline
\end{tabular}
\end{minipage}
\begin{minipage}[t]{0.4\textwidth}
\vspace{0pt}
\begin{tabular}{|cll|}
  \hline
  	$L_s^2L_t$ 	& $\beta_c$ & $\chi_{\left| \overline{l_P}\right|}$\\
  \hline
$64^2 8$	&14.005(37) &55.93(88)	\\
$80^2 8$	&13.836(24)	&71.23(136)	\\
$96^2 8$	&13.901(16) &93.98(248)	\\
$112^2 8$	&13.712(16) &106.33(214)	\\
$128^2 8$	&13.736(14) &131.84(390)	\\
$144^2 8$	&13.767(20) &158.18(629)	\\
  \hline
$80^2 10$	&17.096(31) &95.93(174)	\\
$90^2 10$	&16.919(35)	&107.06(234)	\\
$100^2 10$	&16.870(24) &127.64(305)	\\
$110^2 10$	&16.790(53) &137.15(367)	\\
$120^2 10$	&16.731(24) &154.83(462)	\\
$140^2 10$	&16.725(23) &184.45(659)	\\
  \hline
$72^2 12$	&20.648(62)	&107.12(158)	\\
$84^2 12$	&20.202(66) &122.92(199)	\\
$96^2 12$	&20.098(52) &144.13(252)	\\
$120^2 12$	&19.797(29) &190.28(414)	\\
$144^2 12$	&19.757(27) &236.11(603)	\\
  \hline
\end{tabular}
\end{minipage}
	\caption{$\beta_c$ and $\chi_{\left| \overline{l_P}\right|}$ in $SO(4)$ for $L_t\geq6$ on various volumes.}
	\label{phase:tab:so4susceptibilitydatapart2}
\end{table}

\begin{table}
\centering
\begin{tabular}{|cllc|}
  \hline
  	$L_t$ 	& $\beta_c(V\rightarrow\infty)$ & $L_s$ range & $\bar{\chi}^2_\text{dof}$\\
  \hline
2	&6.4891(3) 	&$L_s \geq 20$	&1.19\\
3	&7.573(3)	&$L_s \geq 32$	&1.44\\
4	&8.573(10) 	&$L_s \geq 40$	&1.24\\
6	&10.781(16) &$L_s \geq 48$	&2.77\\
7	&11.980(24) &$L_s \geq 42$	&6.49\\
8	&13.504(34) &$L_s \geq 64$	&13.18\\
10	&16.321(80) &$L_s \geq 90$	&1.19\\
12	&19.090(99) &$L_s \geq 84$	&3.13\\
  \hline
\end{tabular}
	\caption{Infinite volume limit of $\beta_c$, range of volumes used and $\bar{\chi}^2_\text{dof}$ of extrapolation, for $SO(4)$.}
	\label{phase:tab:so4infinitevolume}
\end{table}

\begin{table}
\centering
\begin{minipage}[t]{0.4\textwidth}
\vspace{0pt}
\begin{tabular}{|cll|}
  \hline
  	$L_s^2L_t$ 	& $\beta_c$ & $\chi_{\left| \overline{l_P}\right|}$\\
  \hline
$16^2 2$	&10.380(1) 	&10.75(3)	\\
$18^2 2$	&10.378(1)	&12.82(4)	\\
$20^2 2$	&10.378(1) 	&14.74(7)	\\
$22^2 2$	&10.376(1) 	&16.83(9)	\\
$24^2 2$	&10.377(2) 	&18.53(12)	\\
  \hline
$26^2 3$	&12.058(3) 	&13.41(14)	\\
$28^2 3$	&12.054(3)	&14.07(15)	\\
$30^2 3$	&12.049(3) 	&14.76(12)	\\
$32^2 3$	&12.053(4) 	&15.28(15)	\\
$34^2 3$	&12.049(4) 	&16.02(25)	\\
  \hline  
$32^2 4$	&13.964(10)	&16.42(12)	\\
$36^2 4$	&13.964(9) 	&18.90(13)	\\
$40^2 4$	&13.955(7) 	&21.27(18)	\\
$48^2 4$	&13.962(12)	&24.13(27)	\\
  \hline
$40^2 5$	&16.316(14) &20.75(14)	\\
$44^2 5$	&16.342(8)	&24.15(23)	\\
$50^2 5$	&16.300(13) &26.37(20)	\\
$54^2 5$	&16.295(26) &27.87(18)	\\
$60^2 5$	&16.265(8) 	&30.58(29)	\\
$70^2 5$	&16.290(6) 	&35.40(40)	\\
  \hline    
\end{tabular}
\end{minipage}
\begin{minipage}[t]{0.4\textwidth}
\vspace{0pt}
\begin{tabular}{|cll|}
  \hline
  	$L_s^2L_t$ 	& $\beta_c$ & $\chi_{\left| \overline{l_P}\right|}$\\
  \hline  
$42^2 6$	&19.017(34) &28.35(49)	\\
$48^2 6$	&18.939(61)	&31.74(41)	\\
$54^2 6$	&18.965(19) &36.39(49)	\\
$60^2 6$	&18.930(18) &40.39(76)	\\
  \hline
$56^2 7$	&21.715(17) &46.57(67)	\\
$60^2 7$	&21.663(16)	&48.70(77)	\\
$64^2 7$	&21.592(15) &53.02(81)	\\
$68^2 7$	&21.595(15) &55.50(90)	\\
$72^2 7$	&21.554(11) &58.01(99)	\\
$84^2 7$	&21.564(17) &69.48(159) \\
  \hline
$64^2 8$	&24.329(24) &63.20(102)	\\
$72^2 8$	&24.294(18)	&72.14(124)	\\
$80^2 8$	&24.255(14) &80.30(154)	\\
$88^2 8$	&24.255(17) &85.77(180)	\\
$96^2 8$	&24.182(18) &83.53(197)	\\
  \hline
$80^2 10$	&29.730(19) &109.84(163)	\\
$90^2 10$	&29.621(26)	&116.99(200)	\\
$100^2 10$	&29.521(19) &130.84(236)	\\
$110^2 10$	&29.519(19) &141.11(272)	\\
$120^2 10$	&29.517(22) &159.53(391)	\\
  \hline
\end{tabular}
\end{minipage}
	\caption{$\beta_c(V)$ and $\chi_{\left| \overline{l_P}\right|}(V)$ for $SO(5)$.}
	\label{phase:tab:so5susceptibility}
\end{table}

\begin{table}
\centering
\begin{tabular}{|cllc|}
  \hline
  	$L_t$ 	& $\beta_c(V\rightarrow\infty)$ & $L_s$ range & $\bar{\chi}^2_\text{dof}$\\
  \hline
2	&10.368(3) 	&$L_s \geq 16$	&0.74\\
3	&12.021(16) &$L_s \geq 26$	&0.49\\
4	&13.944(36) &$L_s \geq 32$	&0.32\\
5	&16.180(13) &$L_s \geq 40$	&4.56\\
6	&18.603(55) &$L_s \geq 42$	&1.29\\
7	&21.192(52) &$L_s \geq 56$	&4.63\\
8	&23.938(63) &$L_s \geq 64$	&1.39\\
10	&29.500(157)&$L_s \geq 100$	&0.00\\
  \hline
\end{tabular}
	\caption{Infinite volume limit of $\beta_c$, range of volumes used and $\bar{\chi}^2_\text{dof}$ of extrapolation, for $SO(5)$.}
	\label{phase:tab:so5infinitevolume}
\end{table}

\begin{table}
\centering
\begin{minipage}[t]{0.4\textwidth}
\vspace{0pt}
\begin{tabular}{|cll|}
  \hline
  	$L_s^2L_t$ 	& $\beta_c$ & $\chi_{\left| \overline{l_P}\right|}$\\
  \hline
$8^2 2$		&15.175(2) 	&3.191(8)	\\
$10^2 2$	&15.185(1) 	&4.972(8)	\\
$12^2 2$	&15.192(1) 	&7.151(12)	\\
  \hline
$12^2 3$	&17.810(9) 	&3.94(1)	\\
$16^2 3$	&17.793(4)	&6.22(2)	\\
$20^2 3$	&17.821(4) 	&9.18(3)	\\
$24^2 3$	&17.831(4) 	&12.34(5)	\\
$28^2 3$	&17.833(3) 	&15.79(8)	\\
$32^2 3$	&17.839(3) 	&19.53(13)	\\
  \hline
$28^2 4$	&21.295(6) 	&12.07(5)	\\
$32^2 4$	&21.358(8)	&15.56(12)	\\
$36^2 4$	&21.356(8) 	&18.58(9)	\\
$40^2 4$	&21.352(4) 	&22.23(13)	\\
$44^2 4$	&21.370(6) 	&25.85(15)	\\
  \hline
\end{tabular}
\end{minipage}
\begin{minipage}[t]{0.4\textwidth}
\vspace{0pt}
\begin{tabular}{|cll|}
  \hline
  	$L_s^2L_t$ 	& $\beta_c$ & $\chi_{\left| \overline{l_P}\right|}$\\
  \hline
$28^2 5$	&25.479(24) &14.25(8)	\\
$32^2 5$	&25.496(16)	&17.77(12)	\\
$40^2 5$	&25.501(14) &25.25(18)	\\
$48^2 5$	&25.549(14) &33.99(38)	\\
$56^2 5$	&25.577(11) &44.40(41)	\\
$60^2 5$	&25.589(10) &48.77(68)	\\
  \hline
$42^2 6$	&29.781(26) &31.45(18)	\\
$48^2 6$	&29.727(18)	&38.65(30)	\\
$54^2 6$	&29.791(33) &47.09(75)	\\
$60^2 6$	&29.796(31) &55.66(75)	\\
$66^2 6$	&29.819(9) 	&65.52(89)	\\
  \hline
  
$44^2 7$	&34.031(37) &38.73(30)	\\
$50^2 7$	&33.907(25)	&46.75(48)	\\  
$56^2 7$	&34.078(49) &57.24(96)	\\
$60^2 7$	&34.042(36)	&62.06(65)	\\
$64^2 7$	&34.021(19) &69.71(85)	\\
$68^2 7$	&34.009(25) &75.83(107)	\\
$72^2 7$	&34.092(23) &83.87(143)	\\
  \hline
\end{tabular}
\end{minipage}
	\caption{$\beta_c(V)$ and $\chi_{\left| \overline{l_P}\right|}(V)$ for $SO(6)$.}
	\label{phase:tab:so6susceptibility}
\end{table}

\begin{table}
\centering
\begin{tabular}{|cllc|}
  \hline
  	$L_t$ 	& $\beta_c(V\rightarrow\infty)$ & $L_s$ range & $\bar{\chi}^2_\text{dof}$\\
  \hline
2	&15.205(3) &$L_s \geq 8$	&0.62\\
3	&17.854(3) &$L_s \geq 12$	&0.83\\
4	&21.399(6) &$L_s \geq 28$	&6.42\\
5	&25.613(11) &$L_s \geq 28$	&2.25\\
6	&29.872(21) &$L_s \geq 42$	&2.79\\
7	&34.113(32) &$L_s \geq 44$	&4.44\\
  \hline
\end{tabular}
	\caption{Infinite volume limit of $\beta_c$, range of volumes used and $\bar{\chi}^2_\text{dof}$ of extrapolation, for $SO(6)$.}
	\label{phase:tab:so6infinitevolume}
\end{table}

\begin{table}
\centering
\begin{minipage}[t]{0.4\textwidth}
\vspace{0pt}
\begin{tabular}{|cll|}
  \hline
  	$L_s^2L_t$ 	& $\beta_c$ & $\chi_{\left| \overline{l_P}\right|}$\\
  \hline
$8^2 2$		&20.963(3) 	&3.347(3)	\\
$10^2 2$	&20.960(3)	&5.381(10)	\\
$12^2 2$	&20.953(3) 	&7.916(14)	\\
  \hline
$12^2 3$	&25.148(29) &4.00(3)	\\
$16^2 3$	&25.022(14)	&6.60(4)	\\
$20^2 3$	&24.982(13) &9.96(8)	\\
$24^2 3$	&24.988(5) 	&14.27(7)	\\
$28^2 3$	&25.011(7) 	&19.40(14)	\\
  \hline
$24^2 4$	&30.721(24) &12.65(12)	\\
$32^2 4$	&30.714(21)	&21.04(30)	\\
$40^2 4$	&30.721(7) 	&31.62(27)	\\
$48^2 4$	&30.727(6) 	&44.05(43)	\\
$56^2 4$	&30.726(8) 	&58.67(67)	\\
  \hline
  \end{tabular}
\end{minipage}
\begin{minipage}[t]{0.4\textwidth}
\vspace{0pt}
\begin{tabular}{|cll|}
  \hline
  	$L_s^2L_t$ 	& $\beta_c$ & $\chi_{\left| \overline{l_P}\right|}$\\
  \hline
$32^2 5$	&36.909(20) &23.75(21)	\\
$40^2 5$	&36.885(19)	&35.40(43)	\\
$48^2 5$	&36.895(24) &50.05(80)	\\
$56^2 5$	&36.930(22) &66.59(110)	\\
$64^2 5$	&36.909(19) &83.37(151)	\\
  \hline  
$52^2 6$	&43.088(25) &62.57(104)	\\
$56^2 6$	&43.089(10)	&72.92(93)	\\  
$60^2 6$	&43.164(14) &83.65(127)	\\
$64^2 6$	&43.129(17)	&94.02(115)	\\
  \hline
\end{tabular}
\end{minipage}
	\caption{$\beta_c(V)$ and $\chi_{\left| \overline{l_P}\right|}(V)$ for $SO(7)$.}
	\label{phase:tab:so7susceptibility}
\end{table}

\begin{table}
\centering
\begin{tabular}{|cllc|}
  \hline
  	$L_t$ 	& $\beta_c(V\rightarrow\infty)$ & $L_s$ range & $\bar{\chi}^2_\text{dof}$\\
  \hline
2	&20.947(6) 	&$L_s \geq 8$	&0.82\\
3	&24.992(11) &$L_s \geq 12$	&5.46\\
4	&30.729(9) 	&$L_s \geq 24$	&0.14\\
5	&36.913(20) &$L_s \geq 32$	&0.81\\
6	&43.311(62) &$L_s \geq 52$	&5.00\\
  \hline
\end{tabular}
	\caption{Infinite volume limit of $\beta_c$, range of volumes used and $\bar{\chi}^2_\text{dof}$ of extrapolation, for $SO(7)$.}
	\label{phase:tab:so7infinitevolume}
\end{table}

\begin{table}
\centering
\begin{minipage}[t]{0.4\textwidth}
\vspace{0pt}
\begin{tabular}{|cll|}
  \hline
  	$L_s^2L_t$ 	& $\beta_c$ & $\chi_{\left| \overline{l_P}\right|}$\\
  \hline
$8^2 2$		&27.583(4) 	&3.241(4)	\\
$10^2 2$	&27.594(3) 	&5.279(6)	\\
$12^2 2$	&27.605(3)	&7.851(7)	\\
  \hline
$16^2 3$	&33.488(22) &6.29(6)	\\
$18^2 3$	&33.503(17)	&8.05(7)	\\
$20^2 3$	&33.468(13) &9.68(5)	\\
$24^2 3$	&33.521(8) 	&14.54(7)	\\
  \hline
$24^2 4$	&41.573(14) &13.20(6)	\\
$28^2 4$	&41.600(18) &17.88(16)	\\
$32^2 4$	&41.631(14)	&23.71(17)	\\
$40^2 4$	&41.686(14) &37.98(28)	\\
  \hline
\end{tabular}
\end{minipage}
\begin{minipage}[t]{0.4\textwidth}
\vspace{0pt}
\begin{tabular}{|cll|}
  \hline
  	$L_s^2L_t$ 	& $\beta_c$ & $\chi_{\left| \overline{l_P}\right|}$\\
  \hline
$32^2 5$	&50.073(23) &25.70(27)	\\
$40^2 5$	&50.163(21)	&40.21(40)	\\
$48^2 5$	&50.178(37) &58.72(72)	\\
$56^2 5$	&50.247(22) &81.93(90)	\\
  \hline
$42^2 6$	&58.560(18) &46.27(34)	\\
$48^2 6$	&58.742(24)	&63.38(78)	\\  
$54^2 6$	&58.703(19) &79.59(60)	\\
$60^2 6$	&58.758(12)	&101.15(99)	\\
  \hline
\end{tabular}
\end{minipage}
	\caption{$\beta_c(V)$ and $\chi_{\left| \overline{l_P}\right|}(V)$ for $SO(8)$.}
	\label{phase:tab:so8susceptibility}
\end{table}

\begin{table}
\centering
\begin{tabular}{|cllc|}
  \hline
  	$L_t$ 	& $\beta_c(V\rightarrow\infty)$ & $L_s$ range & $\bar{\chi}^2_\text{dof}$\\
  \hline
2	&27.622(5) &$L_s \geq 8$	&0.66\\
3	&33.547(21) &$L_s \geq 16$	&4.06\\
4	&41.769(32) &$L_s \geq 24$	&0.22\\
5	&50.319(32) &$L_s \geq 32$	&0.46\\
6	&58.935(29) &$L_s \geq 42$	&6.65\\
  \hline
\end{tabular}
	\caption{Infinite volume limit of $\beta_c$, range of volumes used and $\bar{\chi}^2_\text{dof}$ of extrapolation, for $SO(8)$.}
	\label{phase:tab:so8infinitevolume}
\end{table}

\begin{table}
\centering
\begin{minipage}[t]{0.4\textwidth}
\vspace{0pt}
\begin{tabular}{|cll|}
  \hline
  	$L_s^2L_t$ 	& $\beta_c$ & $\chi_{\left| \overline{l_P}\right|}$\\
  \hline
$16^2 3$	&43.443(9) 	&6.79(2)	\\
$18^2 3$	&43.425(6)	&8.64(2)	\\
$20^2 3$	&43.427(7) 	&10.80(4)	\\
$22^2 3$	&43.431(7) 	&13.26(5)	\\
$24^2 3$	&43.460(7) 	&16.16(5)	\\
$26^2 3$	&43.457(6) 	&19.17(6)	\\
$30^2 3$	&43.424(7) 	&25.59(9)	\\
$36^2 3$	&43.433(12) &37.59(26)	\\
$42^2 3$	&43.399(8) 	&51.23(30)	\\
  \hline
\end{tabular}
\end{minipage}
\begin{minipage}[t]{0.4\textwidth}
\vspace{0pt}
\begin{tabular}{|cll|}
  \hline
  	$L_s^2L_t$ 	& $\beta_c$ & $\chi_{\left| \overline{l_P}\right|}$\\
  \hline
$20^2 4$	&54.449(28) &10.96(10)	\\
$24^2 4$	&54.375(28)	&15.93(14)	\\
$28^2 4$	&54.393(19) &22.01(16)	\\
$32^2 4$	&54.505(17) &30.22(18)	\\
$40^2 4$	&54.464(14) &48.16(26)	\\
$48^2 4$	&54.434(12) &70.28(55)	\\
  \hline
$24^2 5$	&65.634(38) &16.87(9)	\\
$28^2 5$	&65.654(46)	&23.34(16)	\\
$32^2 5$	&65.661(38) &31.17(22)	\\
$36^2 5$	&65.674(15) &40.16(22)	\\
$40^2 5$	&65.648(20) &50.10(36)	\\
$48^2 5$	&65.760(17) &75.33(53)	\\
  \hline
\end{tabular}
\end{minipage}
	\caption{$\beta_c(V)$ and $\chi_{\left| \overline{l_P}\right|}(V)$ for $SO(9)$.}
	\label{phase:tab:so9susceptibility}
\end{table}

\begin{table}
\centering
\begin{tabular}{|cllc|}
  \hline
  	$L_t$ 	& $\beta_c(V\rightarrow\infty)$ & $L_s$ range & $\bar{\chi}^2_\text{dof}$\\
  \hline
3	&43.450(7) 	&$L_s \geq 16$	&6.10\\
4	&54.457(13) &$L_s \geq 20$	&6.78\\
5	&65.678(32) &$L_s \geq 24$	&0.47\\
  \hline
\end{tabular}
	\caption{Infinite volume limit of $\beta_c$, range of volumes used and $\bar{\chi}^2_\text{dof}$ of extrapolation, for $SO(9)$.}
	\label{phase:tab:so9infinitevolume}
\end{table}

\begin{table}
\centering
\begin{minipage}[t]{0.4\textwidth}
\vspace{0pt}
\begin{tabular}{|cll|}
  \hline
  	$L_s^2L_t$ 	& $\beta_c$ & $\chi_{\left| \overline{l_P}\right|}$\\
  \hline
$6^2 2$	&63.57(1) 	&1.593(1)	\\
$7^2 2$	&63.59(1)	&2.248(1)	\\
$8^2 2$	&63.58(1) 	&3.014(3)	\\
  \hline
$8^2 3$		&80.72(1) 	&1.620(3)	\\
$10^2 3$	&80.83(1) 	&2.640(8)	\\
$12^2 3$	&80.96(1) 	&4.043(8)	\\
$14^2 3$	&81.05(1) 	&5.800(16)	\\
$16^2 3$	&81.12(1) 	&7.929(17)	\\
  \hline
\end{tabular}
\end{minipage}
\begin{minipage}[t]{0.4\textwidth}
\vspace{0pt}
\begin{tabular}{|cll|}
  \hline
  	$L_s^2L_t$ 	& $\beta_c$ & $\chi_{\left| \overline{l_P}\right|}$\\
  \hline
$12^2 4$	&101.81(5) 	&3.995(25)	\\
$16^2 4$	&102.11(8) 	&7.619(84)	\\
$20^2 4$	&102.19(4) 	&12.739(71)	\\
$24^2 4$	&102.28(4)	&19.393(88)	\\
  \hline
$16^2 5$	&123.03(10) &7.709(60)	\\
$20^2 5$	&123.26(8)	&12.988(104)	\\
$24^2 5$	&123.52(4) 	&19.769(137)	\\
$28^2 5$	&123.62(6)	&27.835(284)	\\
  \hline
\end{tabular}
\end{minipage}
	\caption{$\beta_c(V)$ and $\chi_{\left| \overline{l_P}\right|}(V)$ for $SO(12)$.}
	\label{phase:tab:so12susceptibility}
\end{table}

\begin{table}
\centering
\begin{tabular}{|cllc|}
  \hline
  	$L_t$ 	& $\beta_c(V\rightarrow\infty)$ & $L_s$ range & $\bar{\chi}^2_\text{dof}$\\
  \hline
2	&63.610(14) &$L_s \geq 6$	&3.35\\  
3	&81.299(17) &$L_s \geq 8$	&0.70\\
4	&102.424(45) &$L_s \geq 12$	&0.16\\
5	&124.011(15) &$L_s \geq 16$	&0.34\\
  \hline
\end{tabular}
	\caption{Infinite volume limit of $\beta_c$, range of volumes used and $\bar{\chi}^2_\text{dof}$ of extrapolation, for $SO(12)$.}
	\label{phase:tab:so12infinitevolume}
\end{table}

\begin{table}
\centering
\begin{minipage}[t]{0.4\textwidth}
\vspace{0pt}
\begin{tabular}{|cll|}
  \hline
  	$L_s^2L_t$ 	& $\beta_c$ & $\chi_{\left| \overline{l_P}\right|}$\\
  \hline
$4^2 2$	&114.85(2) 	&0.595(1)	\\
$5^2 2$	&114.82(2)	&0.998(2)	\\
$6^2 2$	&114.84(2) 	&1.512(3)	\\
  \hline
$6^2 3$		&149.17(2) 	&0.940(2)	\\
$8^2 3$		&149.32(2) 	&1.832(3)	\\
$10^2 3$	&149.58(3) 	&3.141(4)	\\
$12^2 3$	&149.76(2) 	&4.839(10)	\\
$14^2 3$	&149.89(3) 	&6.897(20)	\\
  \hline
\end{tabular}
\end{minipage}
\begin{minipage}[t]{0.4\textwidth}
\vspace{0pt}
\begin{tabular}{|cll|}
  \hline
  	$L_s^2L_t$ 	& $\beta_c$ & $\chi_{\left| \overline{l_P}\right|}$\\
  \hline
$6^2 4$		&192.07(26)	&1.016(7)	\\
$8^2 4$		&189.04(32)	&1.804(25)	\\
$10^2 4$	&188.91(17) &3.011(42)	\\
$12^2 4$	&189.13(11)	&4.617(47)	\\
$14^2 4$	&189.33(11) &6.725(74)	\\
$16^2 4$	&189.55(6) 	&9.221(40)	\\
  \hline
$8^2 5$		&230.62(44)	&1.972(18)	\\
$10^2 5$	&229.42(56)	&3.013(44)	\\
$12^2 5$	&228.60(5)	&4.591(14)	\\
$14^2 5$	&228.77(6)	&6.572(18)	\\
$16^2 5$	&229.01(7) 	&9.036(31)	\\
$20^2 5$	&229.51(12)	&15.429(88)	\\
  \hline
\end{tabular}
\end{minipage}
	\caption{$\beta_c(V)$ and $\chi_{\left| \overline{l_P}\right|}(V)$ for $SO(16)$.}
	\label{phase:tab:so16susceptibility}
\end{table}

\begin{table}
\centering
\begin{tabular}{|cllc|}
  \hline
  	$L_t$ 	& $\beta_c(V\rightarrow\infty)$ & $L_s$ range & $\bar{\chi}^2_\text{dof}$\\
  \hline
2	&114.824(35) &$L_s \geq 4$	&0.66\\  
3	&150.128(33) &$L_s \geq 8$	&0.83\\
4	&189.975(14) &$L_s \geq 12$	&0.19\\
5	&230.096(212) &$L_s \geq 14$	&1.06\\
  \hline
\end{tabular}
	\caption{Infinite volume limit of $\beta_c$, range of volumes used and $\bar{\chi}^2_\text{dof}$ of extrapolation, for $SO(16)$.}
	\label{phase:tab:so16infinitevolume}
\end{table}

\clearpage

\begin{table}
\centering
\begin{tabular}{|cclc|}
  \hline
  	$N$ 	& $L_s^2L_t$ & Bulk transition & Weak coupling region\\
  \hline
$4$	&$20^2 24$	&$\beta\in[9.1,10.2]$ 	&$L_t\geq 6$	\\
$5$	&$12^2 24$	&$\beta\in[13.5,15.4]$ 	&$L_t\geq 5$	\\
$6$	&$12^2 24$	&$\beta\in[18.0,21.3]$ 	&$L_t\geq 5$	\\
$7$	&$8^2 24$	&$\beta\in[23.5,28.0]$ 	&$L_t\geq 4$	\\
$8$	&$8^2 24$	&$\beta\in[31,35]$ 	&$L_t\geq 4$	\\
$9$	&$4^2 24$	&$\beta\in[37,42]$ 	&$L_t\geq 3$	\\
$12$&$4^2 24$	&$\beta\in[65,73]$ 	&$L_t\geq 3$	\\
$16$&$2^2 24$	&$\beta\in[111,124]$ 	&$L_t\geq 3$	\\
  \hline
\end{tabular}
	\caption{Location of the bulk transition on volumes $L_s^2L_t$, and consequent range of $L_t$ for which $\beta_c$ is on weak coupling side.}
	\label{phase:tab:bulktransition}
\end{table}

\begin{table}
\centering
\begin{tabular}{|clllllc|}
  \hline
  	$L_t$ 	& $\beta_c(V\rightarrow\infty)$ & $a\surd\sigma$ & $T_c/\surd\sigma$ & $\beta_I(V\rightarrow\infty)$ & $T_c/(g^2N)$ & Coupling\\
  \hline
2	&6.4891(3) 	&0.6208(3)	&0.8054(3) 	&3.7234(5)	&0.05818(1) & \multirow{3}{*}{Strong}\\
3	&7.573(3)	&0.3970(7)	&0.8397(14) &5.169(3)	&0.05384(3) & \\
4	&8.573(10) 	&0.2936(10)	&0.8515(31) &6.290(11)	&0.04914(8) & \\
\hline
6	&10.781(16) &0.2146(17)	&0.7766(60) &8.590(16)	&0.04474(8) & \multirow{5}{*}{Weak}\\
7	&11.980(24) &0.1845(11)	&0.7743(44) &9.816(24)	&0.04382(11) & \\
8	&13.504(34) &0.1609(12)	&0.7770(60) &11.364(34)	&0.04439(13) & \\
10	&16.322(80) &0.1206(13)	&0.7718(81) &14.213(81)	&0.04442(25) & \\
12	&19.090(99) &0.1083(15)	&0.7692(106)&17.099(76)	&0.04453(20) & \\
  \hline
\end{tabular}
	\caption{$SO(4)$ critical temperature in units of the string tension, $T_c/\surd\sigma$, and in units of the (mean field improved) 't Hooft coupling, $T_c/(g^2N)$, evaluated at $\beta_c(V\rightarrow\infty)$.}
	\label{phase:tab:so4continuum}
\end{table}

\begin{table}
\centering
\begin{tabular}{|clllllc|}
  \hline
  	$L_t$ 	& $\beta_c(V\rightarrow\infty)$ & $a\surd\sigma$ & $T_c/\surd\sigma$ & $\beta_I(V\rightarrow\infty)$ & $T_c/(g^2N)$ & Coupling\\
  \hline
2	&10.368(3) 	&0.6447(14)	&0.7756(17) &5.811(5)	&0.05811(5) & \multirow{3}{*}{Strong}\\
3	&12.021(16)	&0.4248(19)	&0.7847(36) &8.044(19)	&0.05363(13) & \\
4	&13.944(36) &0.3214(15)	&0.7778(36) &10.162(39)	&0.05081(19) & \\
\hline  
5	&16.180(13) &0.2593(15)	&0.7714(44) &12.497(13)	&0.04999(5) & \multirow{5}{*}{Weak}\\
6	&18.603(55) &0.2190(12)	&0.7611(43)	&14.985(56)	&0.04995(19) &\\
7	&21.192(52) &0.1877(11)	&0.7613(45)	&17.620(53)	&0.05034(15) &\\
8	&23.938(63) &0.1624(9)	&0.7698(41)	&10.401(64)	&0.05100(16) &\\
10	&29.500(157) &0.1281(10)&0.7801(62)	&26.010(158)&0.05202(32) &\\
  \hline
\end{tabular}
	\caption{$SO(5)$ critical temperature in units of the string tension, $T_c/\surd\sigma$, and in units of the (mean field improved) 't Hooft coupling, $T_c/(g^2N)$, evaluated at $\beta_c(V\rightarrow\infty)$.}
	\label{phase:tab:so5continuum}
\end{table}

\begin{table}
\centering
\begin{tabular}{|clllllc|}
  \hline
  	$L_t$ 	& $\beta_c(V\rightarrow\infty)$ & $a\surd\sigma$ & $T_c/\surd\sigma$ & $\beta_I(V\rightarrow\infty)$ & $T_c/(g^2N)$ & Coupling\\
  \hline
2	&15.205(3) &0.6535(13)	&0.7651(15) &8.460(4)	&0.05875(3)	& \multirow{3}{*}{Strong}\\
3	&17.854(3) &0.4201(7)	&0.7935(13)	&11.964(4)	&0.05539(2)\\
4	&21.399(6) &0.3106(10)	&0.8050(27)	&15.784(6)	&0.05480(2)\\
  \hline
5	&25.613(11) &0.2501(7)	&0.7996(22) &20.147(11)	&0.05596(3)& \multirow{3}{*}{Weak}\\
6	&29.872(21) &0.2077(3)	&0.8024(14)	&24.497(21)	&0.05670(5)&\\
7	&34.113(32) &0.1774(4)	&0.8053(19)	&28.798(32)	&0.05714(6)&\\
  \hline
\end{tabular}
	\caption{$SO(6)$ critical temperature in units of the string tension, $T_c/\surd\sigma$, and in units of the (mean field improved) 't Hooft coupling, $T_c/(g^2N)$, evaluated at $\beta_c(V\rightarrow\infty)$.}
	\label{phase:tab:so6continuum}
\end{table}

\begin{table}
\centering
\begin{tabular}{|clllllc|}
  \hline
  	$L_t$ 	& $\beta_c(V\rightarrow\infty)$ & $a\surd\sigma$ & $T_c/\surd\sigma$ & $\beta_I(V\rightarrow\infty)$ & $T_c/(g^2N)$ & Coupling\\
  \hline
2	&20.947(6) 	&0.6571(8)	&0.7610(10)	&11.597(9)	&0.05917(4)	& \multirow{2}{*}{Strong}\\
3	&24.992(11) &0.4181(7)	&0.7972(13)	&16.816(12)	&0.05720(4)\\
  \hline
4	&30.729(9) 	&0.3104(5)	&0.8053(12)	&22.924(10)	&0.05848(2)	& \multirow{3}{*}{Weak}\\
5	&36.913(20) &0.2455(7)	&0.8147(22)	&29.302(20)	&0.05980(4)	&\\
6	&43.311(62) &0.2023(6)	&0.8239(26)	&35.821(63)	&0.06092(11)&\\
  \hline
\end{tabular}
	\caption{$SO(7)$ critical temperature in units of the string tension, $T_c/\surd\sigma$, and in units of the (mean field improved) 't Hooft coupling, $T_c/(g^2N)$, evaluated at $\beta_c(V\rightarrow\infty)$.}
	\label{phase:tab:so7continuum}
\end{table}

\begin{table}
\centering
\begin{tabular}{|clllllc|}
  \hline
  	$L_t$ 	& $\beta_c(V\rightarrow\infty)$ & $a\surd\sigma$ & $T_c/\surd\sigma$ & $\beta_I(V\rightarrow\infty)$ & $T_c/(g^2N)$ & Coupling\\
  \hline
2	&27.622(5) 	&0.6586(8)	&0.7591(9)	&15.266(8)	&0.05963(3)	& \multirow{2}{*}{Strong}\\
3	&33.547(21) &0.4179(7)	&0.7977(14)	&22.715(23)	&0.05915(6)	&\\
  \hline
4	&41.769(32) &0.3051(11)	&0.8193(28)	&31.414(33)	&0.06136(6)	& \multirow{3}{*}{Weak}\\
5	&50.319(32) &0.2415(4)	&0.8281(15)	&40.210(32)	&0.06283(5)	&\\
6	&58.935(29) &0.2003(5)	&0.8320(20)	&48.975(29)	&0.06377(4) &\\
  \hline
\end{tabular}
	\caption{$SO(8)$ critical temperature in units of the string tension, $T_c/\surd\sigma$, and in units of the (mean field improved) 't Hooft coupling, $T_c/(g^2N)$, evaluated at $\beta_c(V\rightarrow\infty)$.}
	\label{phase:tab:so8continuum}
\end{table}

\begin{table}
\centering
\begin{tabular}{|clllllc|}
  \hline
  	$L_t$ 	& $\beta_c(V\rightarrow\infty)$ & $a\surd\sigma$ & $T_c/\surd\sigma$ & $\beta_I(V\rightarrow\infty)$ & $T_c/(g^2N)$ & Coupling\\
  \hline
3	&43.450(7)  &0.4150(2)	&0.8032(4)	&29.586(8)	&0.060877(16)	& \multirow{3}{*}{Weak}\\
4	&54.457(13) &0.3025(4)	&0.8263(11)	&41.190(13)	&0.063564(21)	&\\
5	&65.678(32) &0.2395(3)	&0.8351(12)	&52.714(33)	&0.065079(40)	&\\
  \hline
\end{tabular}
	\caption{$SO(9)$ critical temperature in units of the string tension, $T_c/\surd\sigma$, and in units of the (mean field improved) 't Hooft coupling, $T_c/(g^2N)$, evaluated at $\beta_c(V\rightarrow\infty)$.}
	\label{phase:tab:so9continuum}
\end{table}

\begin{table}
\centering
\begin{tabular}{|clllllc|}
  \hline
  	$L_t$ 	& $\beta_c(V\rightarrow\infty)$ & $a\surd\sigma$ & $T_c/\surd\sigma$ & $\beta_I(V\rightarrow\infty)$ & $T_c/(g^2N)$ & Coupling\\
  \hline
2	&63.610(14)  &0.6600(24)	&0.7576(27)	&35.213(22)	&0.06113(4)&Strong \\  
\hline
3	&81.299(17)  &0.4070(6)		&0.8191(12)	&56.114(18)	&0.064947(21)&\multirow{3}{*}{Weak}\\
4	&102.424(45) &0.2977(6)		&0.8399(18)	&78.245(46)	&0.06792(4)	&\\
5	&124.011(15) &0.2354(12)	&0.8497(43)	&100.359(180)&0.06969(12)&\\
  \hline
\end{tabular}
	\caption{$SO(12)$ critical temperature in units of the string tension, $T_c/\surd\sigma$, and in units of the (mean field improved) 't Hooft coupling, $T_c/(g^2N)$, evaluated at $\beta_c(V\rightarrow\infty)$.}
	\label{phase:tab:so12continuum}
\end{table}

\begin{table}
\centering
\begin{tabular}{|clllllc|}
  \hline
  	$L_t$ 	& $\beta_c(V\rightarrow\infty)$ & $a\surd\sigma$ & $T_c/\surd\sigma$ & $\beta_I(V\rightarrow\infty)$ & $T_c/(g^2N)$ & Coupling\\
  \hline
2	&63.610(14)  &0.6438(25)&0.7766(31)	&64.219(47)	&0.06271(5)	&Strong \\  
\hline
3	&81.299(17)  &0.4003(12)&0.8328(24)	&104.617(35)&0.068110(23)&\multirow{3}{*}{Weak}\\
4	&102.424(45) &0.2925(9)	&0.8547(26)	&146.288(207)&0.07143(10)&\\
5	&124.011(15) &0.2320(7)	&0.8622(28)	&187.090(218)&0.07308(9)\\
  \hline
\end{tabular}
	\caption{$SO(16)$ critical temperature in units of the string tension, $T_c/\surd\sigma$, and in units of the (mean field improved) 't Hooft coupling, $T_c/(g^2N)$, evaluated at $\beta_c(V\rightarrow\infty)$.}
	\label{phase:tab:so16continuum}
\end{table}

\begin{table}
\centering
\begin{tabular}{|c|ll|ll|l|}
  \hline
  $N$ & $T_c/\surd\sigma$ & $\bar{\chi}^2_\text{dof}$
  & $T_c/(g^2N)$ & $\bar{\chi}^2_\text{dof}$ & $T_c/M_{0^+}$ \\
  \hline
3*	&0.7072(80)  &     &0.03236(45) &     &0.2373(33) \\
4	&0.7702(88)  &0.12 &0.04567(43) &2.17 &0.2293(30)	\\
5	&0.7963(114) &0.00 &0.05544(93) &0.05 &0.2244(33)	\\
6	&0.8105(42) &0.16 &0.05996(19) &0.53 &0.2232(14)	\\
7	&0.8351(38) &0.98 &0.06478(18) &4.01 &0.2234(12)	\\
8	&0.8418(39) &0.05 &0.06809(16) &0.00 &0.2224(15)	\\
9	&0.8515(15) &0.30 &0.07043(7)  &0.10 &	\\
12	&0.8642(38) &0.02 &0.07552(14) &0.63 &0.2217(18)	\\
16	&0.8780(38) &0.15 &0.07947(17) &0.85 &0.2220(22)	\\
  \hline
\end{tabular}
\caption{$SO(N)$ continuum limit of the deconfining temperature in units of the string tension, $T_c/\surd\sigma$, of the 't Hooft coupling, $T_c/(g^2N)$, and of the lightest scalar glueball mass, $T_c/M_{0^+}$, with corresponding $\bar{\chi}^2_\text{dof}$ of the fits.  Note that we infer the $SO(3)$ value (*) from the $SU(2)$ value.}
	\label{phase:tab:soncontinuum}
\end{table}

\begin{table}
\centering
\begin{tabular}{|c|ll|ll|ll|}
  \hline
  & \multicolumn{2}{|c|}{$T_c/\surd\sigma$}
  & \multicolumn{2}{|c|}{$T_c/g^2N$}
  & \multicolumn{2}{|c|}{$T_c/M_{0^+}$}  \\
  \hline
  	$N$ 	& $SO(2N)$ & $SU(N)$ 	& $SO(2N)$ & $SU(N)$ 	& $SO(2N)$ & $SU(N)$ \\
  \hline
2	&0.7702(88) &1.1238(88)	&0.0913(9) &0.1998(34) &0.2293(30) &0.2373(19)\\
3	&0.8105(42) &0.9994(40)	&0.1199(4) &0.1904(12) &0.2232(14) &0.2288(10)\\
4	&0.8418(39) &0.9572(39)	&0.1362(3) &0.1884(12) &0.2224(15) &0.2259(11)\\
5	&	    &0.9380(19) &	   &0.1874(10) &	   &0.2233(7)	\\
6	&0.8642(38) &0.9300(48)	&0.1510(3) &0.1873(8)  &0.2217(18) &0.2232(12)\\
8	&0.8780(38) &0.9144(41)	&0.1589(3) &0.1849(10) &0.2220(22) &0.2207(11)	\\
  \hline
\end{tabular}
	\caption{$SO(2N)$ and $SU(N)$ \cite{Jack-Liddle:2008xr} continuum limit of the deconfining temperature in units of the string tension, $T_c/\surd\sigma$, the 't Hooft coupling, $T_c/(g^2N)$, and the lightest scalar glueball mass, $T_c/M_{0^+}$.} 
	\label{phase:tab:sonsun}
\end{table}

\clearpage

\section{Figures}
\label{app:figures}

\vspace{1cm}

\end{document}